\documentclass[12pt]{iopart}

\usepackage{iopams}
\usepackage{graphicx}
\usepackage{bm}
\usepackage{hyperref}
\usepackage{graphics}
\usepackage{slashed}
\usepackage{amssymb}
\usepackage{url}
\usepackage[usenames,dvipsnames,svgnames,table]{xcolor}
\usepackage[T1]{fontenc}

\newcommand{\fracc}[2]{\frac{\textstyle{#1}}{\textstyle{#2}}}

\makeatletter
\newcommand{\tpitchfork}{%
  \vbox{
    \baselineskip\z@skip
    \lineskip-.52ex
    \lineskiplimit\maxdimen
    \m@th
    \ialign{##\crcr\hidewidth\smash{$-$}\hidewidth\crcr$\pitchfork$\crcr}
  }%
}
\makeatother

\usepackage[normalem]{ulem}
\usepackage{color}

\begin{document}

\title{Qualitative analysis of a quasi-magnetic universe}
\author{Alan G. Cesar${}^{1}$, Mario Novello${}^{1}$, Eduardo Bittencourt${}^{2}$\footnote{Corresponding author} and Fernando A. Franco${}^{2}$}
\address{${}^{1}$Centro Brasileiro de Pesquisas F\'isicas, R. Dr. Xavier Sigaud, 150, Botafogo, Rio de Janeiro/RJ, Brazil}
\address{${}^{2}$Universidade Federal de Itajub\'a, Av. BPS 1303, Pinheirinho, Itajub\'a-MG, 37500-903, Brazil}
\ead{alangois@cbpf.br, novello@cbpf.br, bittencourt@unifei.edu.br, fernandoaf@unifei.edu.br}

\date{\today}

\begin{abstract}
We investigate the cosmological dynamics induced by nonlinear electrodynamics in a homogeneous and isotropic universe, focusing on the role of primordial electromagnetic fields with random spatial orientations. Building upon a generalization of the Tolman–Ehrenfest averaging procedure, we derive a modified energy-momentum tensor consistent with the spacetime symmetries, incorporating the influence of the dual invariant $G$ and its statistical contributions. A specific nonlinear electrodynamics model with quadratic corrections to Maxwell’s Lagrangian is considered, giving rise to what we define as a quasi-magnetic universe, interpolating between purely magnetic and statistically null field configurations. We analyze the resulting cosmological dynamics through qualitative methods. By casting the equations into autonomous dynamical systems, we identify the equilibrium points, determine their stability, and study the behavior of solutions under various spatial curvatures. Our findings reveal the existence of bouncing and cyclic solutions, regions where energy conditions are violated, and scenarios of accelerated expansion. Special attention is given to two limiting cases, both of which exhibit qualitatively distinct phase portraits and energy-condition behavior. This work provides a comprehensive framework for understanding the influence of nonlinear electromagnetic fields in the early universe and opens avenues for exploring their observational consequences.
\end{abstract}

\vspace{2pc}

\maketitle

\section{Introduction}
Despite its remarkable success in explaining a wide range of cosmological observations--such as the cosmic microwave background (CMB), large-scale structure, and the apparent accelerated expansion of the universe--the $\Lambda$CDM cosmological model faces significant challenges when confronted with different datasets \cite{Ellis2018}. Some of these issues are recent, such as uncertainties surrounding the nature of dark matter and dark energy \cite{Giare2025,des2025}, as well as the persistent Hubble tension (see Refs.\ \cite{Valentino2021,Hu2023} and references therein). Others are long-standing, including the difficulties posed by the Big Bang singularity, such as the horizon and flatness problems. 

These challenges underscore the necessity for further exploration and refinement of cosmological models, prompting proposals that extend beyond the $\Lambda$CDM framework. For example, in bouncing \cite{Novello2008,Brandenberger2017} or cyclic \cite{Steinhardt2002,Baum2007,Novello2009,Medeiros2012} cosmological scenarios, the homogeneity and isotropy of spacetime are preserved, while the initial singularity and its associated problems are avoided, often with the incorporation of quantum mechanical principles \cite{Sakellariadou2017,Pintoneto2012,Bishop2023}. In contrast, inhomogeneous cosmological models \cite{Krasinski1997,Bolejko2017,Bittencourt2021,Bittencourt2022} deviate from the Cosmological Principle, resulting in a more complex and diverse framework with additional degrees of freedom, albeit at the cost of increased technical complexity. Furthermore, modifications to the Einstein field equations have paved the way for a wide array of alternative theories, such as $f(R)$ \cite{Sotiriou2010,Nojiri2006,Nojiri2017,Nojiri2011}, $f(T)$ \cite{Capozziello2011,Cai2016,Mishra2025a,Mishra2025b}, and $f(Q)$ gravity \cite{Lazkoz2019,Frusciante2021,Atayde2021,Heisenberg2024}, teleparallel gravity \cite{Bengochea2009}, Horndeski theory \cite{Kobayashi2019}, and others \cite{Clifton2012,Pavlovic2017,Moffat2021,saridakis2021, Nojiri2017b}, offering a rich and varied landscape for theoretical exploration. 

In this paper, we explore an alternative approach by investigating how primordial electromagnetic (EM) fields, governed by a nonlinear equation of motion, influence the early dynamics of the universe. Nonlinear electrodynamics (NLED) emerges in various contexts to address challenges within classical electrodynamics, such as establishing bounds on fields \cite{Born1934}, predicting strong-field corrections \cite{akmansoy,Brevik2025}, vacuum birefringence \cite{heisenberg,denisov,kruglov3}, and the nature of light propagation \cite{Herrera, goulart, Novello2001}. These features give rise to several cosmological effects \cite{Breton2010,Kruglov2015,Ovgun2018,Sorokin2022}, including models with accelerated expansion phases \cite{Kruglov2015b,Novello2007,Novello2004, kruglov2, benaoum}, non-singular models \cite{Novello2008,Novello2009,Medeiros2012,Novello2007,Lorenci2002,Novello2012}, and numerous studies examining their implications for black hole dynamics \cite{canate, bakopoulos, maceda, ruffini,Goulart2024,Rahmatov2025}. 

In contrast to previous works that examined either a purely magnetic universe or individual NLED models in isolation \cite{Bittencourt2014,Herrera,Novello2007,Novello2004,Lorenci2002,GarciaSalcedo2010,Dwivedi2024} this study develops a unified dynamical description. Strictly speaking, it (i) generalizes the Tolman–Ehrenfest averaging procedure to systematically include nonlinear effects, (ii) incorporates a nonvanishing electric field, which significantly enriches the branches of primordial cosmic dynamics, even if current cosmological data suggest it is negligible, and (iii) analyzes the phase-space structure of the resulting autonomous dynamical system, including fixed points, separatrices, and energy-condition boundaries.

This gap in the literature motivates our focus on the quadratic NLED sector. Quadratic terms naturally arise as the leading-order corrections in the Euler–Heisenberg effective action \cite{heisenberg} and in the low-field expansion of Born–Infeld theory \cite{Born1934}, making them a well-motivated starting point for a qualitative dynamical study. To place this work within the broader landscape of modified-cosmology research, we emphasize that our approach is driven by the \emph{matter sector}, namely the backreaction of NLED on a homogeneous and isotropic background, rather than by modifications to the gravitational sector itself. Moreover, our assumptions remain physically reasonable, since the hot and dense phase of the primordial universe could naturally excite nonlinear corrections to electromagnetism.

The paper is organized as follows. In Section 2, we discuss a generalization of the Tolman–Ehrenfest averaging procedure in the presence of NLED within a homogeneous and isotropic background. Section 3 introduces a specific nonlinear EM Lagrangian to analyze the cosmological implications of this procedure, together with a relation between the electric and magnetic fields that renders the system tractable. In Section 4, we apply the qualitative theory of differential equations to explore the full dynamics of the universe in this framework: we first identify the equilibrium points and examine their stability, then impose the Friedmann equation as a constraint and discuss the resulting physical solutions for different spatial curvatures. Finally, we consider special limiting cases and analyze their behavior.

\section{Tolman-Ehrenfest average process revisited}
In cosmology, generic EM fields that satisfy the linear Maxwell equations introduce specific directions in space, thereby breaking the symmetries of homogeneity and isotropy. To restore these symmetries, certain conditions must be imposed on the fields. One well-established method for achieving this is the averaging procedure introduced by Tolman and Ehrenfest \cite{Tolman1930,Tolman1987}. 

When dealing with non-linear electrodynamics, the situation is similar. For a given Lagrangian that depends on both EM invariants, the fluid components of the corresponding energy-momentum tensor are generally not spatially homogeneous and isotropic. Consequently, some form of averaging over the fields must be performed to eliminate the directional dependence in the universe. 

However, in this paper, we argue that the standard formulation needs to be generalized to account for the nonlinearities consistently. The complexity arises from the nontrivial form of the Lagrangian and its derivatives, as well as their dependence on the randomness of the fields. This randomness may introduce an additional contribution to the average field, which must be carefully considered.

In order to describe a spatially homogeneous and isotropic universe, we consider a Friedmann-Lema\^itre-Robertson-Walker (FLRW) geometry, as follows
\begin{equation}
    \label{metric}
    ds^2=dt^2-a^2(t)\left[d\chi^2+f^2(\chi)(d\theta^2+\sin^2(\theta)d\phi^2)\right],
\end{equation}
where $a(t)$ is the scale factor, and $f(\chi)$ is $\chi$, $\sin\chi$ or $\sinh\chi$ depending on whether the curvature of the 3-space is flat, spherical, or hyperbolic, respectively. As a source for this geometry, we consider a nonlinear Lagrangian $\mathcal{L}(F,G)$ depending on both EM invariants, defined as $F=F_{\mu\nu}F^{\mu\nu}$ and $G={}^{*}F^{\mu\nu}F_{\mu\nu}$, where $F_{\mu\nu}$ is the Faraday tensor and $^{*}F_{\mu\nu}=\frac{1}{2}\eta_{\mu\nu}{}^{\alpha\beta}F_{\alpha\beta}$ its algebraic dual, with $\eta_{\mu\nu\alpha\beta}$ representing the totally skew-symmetric Levi-Civita tensor. Thus, the energy-momentum tensor, defined as $T^{\mu\nu}=\frac{2}{\sqrt{-g}}\frac{\delta(\sqrt{-g}\mathcal{L})}{\delta g_{\mu\nu}}$, of the EM field is given by
\begin{equation}
    \label{tensormomentumenergia}
T^{\mu\nu}=-4\mathcal{L}_{F}F^{\mu\alpha}F_{\alpha}^{\nu}-(\mathcal{L}-G\mathcal{L}_G)g^{\mu\nu},
\end{equation}
where $\mathcal{L}_{F}$ and $\mathcal{L}_{G}$ are the partial derivatives of the Lagrangian with respect to the invariants $F$ and $G$, respectively. 

In terms of a given congruence of normalized time-like observers $v^{\mu}$, we can define the 3-space orthogonal to it through the projector $h_{\mu\nu}=g_{\mu\nu}-v_{\mu}v_{\nu}$. With this time and space split, we can decompose the Faraday tensor as
\begin{equation}
    F_{\mu\nu}=E_{\mu}v_{\nu} - E_{\nu}v_{\mu} +\eta_{\mu\nu}{}^{\alpha\beta}v_{\alpha}B_{\beta},
\end{equation}
where $E_{\mu}$ represents the electric field and $B_{\nu}$ stands for the magnetic field both seen by $v^{\mu}$. In terms of these fields, the EM invariants are written as $F=2(B^2-E^2)$ and $G=-4E^{\mu}B_{\mu}$, with $E^2=-E_{\mu}E^{\mu}$ and $B^2=-B_{\mu}B^{\mu}$. Then, we apply the same procedure to decompose $T_{\mu\nu}$ into its irreducible parts, obtaining
\begin{eqnarray}
&&\rho = -\mathcal{L}+\mathcal{L}_{G}G-4\mathcal{L}_{F}E^2,\label{densidadedeenergia}\\[1ex]
&&p = \mathcal{L}-\mathcal{L}_{G}G-\frac{4}{3}\mathcal{L}_{F}(F+E^2),\label{pressãoisotrópica}\\[1ex]
&&q_{\lambda}=-4 \mathcal{L}_{F}\eta_{\lambda\gamma\rho\sigma}v^{\rho}B^{\sigma}E^{\gamma},\label{fluxodecalor}\\[1ex]
&&\pi_{\mu\nu}=4 \mathcal{L}_{F}\left(E_{\mu}E_{\nu} + B_{\mu}B_{\nu}\right) + \frac{4}{3}\mathcal{L}_{F}\left(E^2+B^2\right)h_{\mu\nu},\label{pressãoanisotrópica}
\end{eqnarray}
where $\rho$ is the energy density, $p$ is the isotropic pressure, $q_{\lambda}$ is the heat flow, and $\pi_{\mu\nu}$ is the anisotropic pressure tensor.

In general, the off-diagonal components of $T_{\mu\nu}$ are encoded in the components (\ref{fluxodecalor}) and (\ref{pressãoanisotrópica}). Assuming a metric tensor in the form (\ref{metric}), the isometry group of this geometry imposes conditions on the fluid content such that it must reduce to a perfect fluid, only with $\rho$ and $p$, both depending only on time. The conditions to attend such compatibility can be formulated through spatial averages over a delimited region of space, called here a \textit{cosmological cell}, so that the average tensor has only a dependence on time. The formal definition is
\begin{equation}
    \label{médiadetolman}
    \left<X\right>(t) = \lim_{V\rightarrow V_0}\frac{1}{V}\int_{V} X\sqrt{-g} d^3x,
\end{equation}
where $V_0$ represents the volume of the cell.

By applying this procedure to the fluid components (\ref{densidadedeenergia})-(\ref{pressãoanisotrópica}), and imposing the homogeneous and isotropic symmetry to the average fluid, we get the following compatibility conditions:
\begin{equation}
\label{escolha1}
\left<\mathcal{L}_F B^{\mu}E^{\nu}\right>=0,
\end{equation}
and
\begin{equation}
    \label{escolha2}
    \left<\mathcal{L}_F E^{\mu} E^{\nu}\right>+\left<\mathcal{L}_F B^{\mu}B^{\nu}\right>=-\frac{1}{3}\left<\mathcal{L}_F\right>(E^2+B^2)h^{\mu\nu}.
\end{equation}
Note that this leads precisely to $\left<q_{\lambda}\right>=0$ and $\left<\pi_{\mu\nu}\right>=0$, namely, a perfect fluid \textit{on the average}. Also, we have that the only functional dependence allowed is with respect to time, that is, $E=E(t)$ and $B=B(t)$. 

It should be stressed that the procedure employed here is the standard one used in cosmology. Since the spacetime is homogeneous and isotropic, the underlying probability distribution is in fact uniform. The appearance of the Lagrangian and its derivatives may suggest that we are dealing with a weighted average, but this is not the case. These quantities follow directly from the explicit expression of the energy–momentum tensor (\ref{tensormomentumenergia}). Therefore, the uniqueness and covariance are ensured by the spacetime symmetries, exactly as in the standard Tolman–Ehrenfest prescription. In our case, the average of $T_{\mu\nu}$ is different from the linear case precisely due to a random angular contribution introduced by the invariant $G$ through the Lagrangian and its derivative, while the invariant $F$ is spatially independent in this cosmological scenario. Thus, when averaging over random field orientations, the leading effect of small-scale directional fluctuations on the macroscopic EM tensor is controlled by the sensitivity $\mathcal{L}_{F}$. Analogous comments apply to $\mathcal{L}$ and $\mathcal{L}_{G}$ when averaging the energy density and the pressure. In fact, their averages over the cell are
\begin{eqnarray}
&&\left<\rho\right> = -\left<\mathcal{L}\right>+\left<\mathcal{L}_{G}G\right> - 4E^2\left<\mathcal{L}_{F}\right>,\label{rho_mean}\\[1ex]
&&\left<p\right> = \left<\mathcal{L}\right>-\left<\mathcal{L}_{G}G\right> - \frac{4}{3}(F+E^2)\left<\mathcal{L}_{F}\right>.\label{p_mean}
\end{eqnarray}

Additionally, we also have to guarantee that the equation of motion of the EM field is satisfied on average. Therefore,
\begin{equation}
    \label{equaçãodemovimento}
    \partial_{\nu}\left(\mathcal{L}_F F^{\mu\nu}+\mathcal{L}_{G}{}^{*}F^{\mu\nu}\right)=0
\end{equation}
must be identically satisfied over the cosmological cell. By applying the average over this expression and using the Faraday tensor decomposition, we obtain the constraints
\begin{equation}
\label{eom_mean_const}
\left<\mathcal{L}_F E^{\mu}\right>=-\left<\mathcal{L}_G B^{\mu}\right>,\quad\mbox{and}\quad
\left<\mathcal{L}_G E^{\mu}\right>=\left<\mathcal{L}_F B^{\mu}\right>.
\end{equation}

Although there are specific Lagrangians that work well within the Tolman-Ehrenfest procedure \cite{Lorenci2002,Bittencourt2014}, including the linear Maxwell's theory, we believe this must be the general way to treat the average of nonlinear EM fields in a cosmological scenario. The Lagrangian and its derivative should be included in the average to consider as a weight function. Finally, it is worth mentioning that the standard approach can be recovered straightforwardly when the EM theory is linear. Indeed, considering Maxwell's
Lagrangian $\mathcal{L}=-\frac{1}{4}F$, with derivatives $\mathcal{L}_F=-\frac{1}{4}$ and $\mathcal{L}_G=0$, all averages of the form $\langle \mathcal{L}_{F,G} X\rangle$ reduce to $\mathcal{L}_{F,G}\langle X\rangle$, and the averaged energy-momentum tensor recovers precisely the isotropized Maxwell form used in classical cosmology, while the generalized averages (\ref{escolha1}), (\ref{escolha2}) and (\ref{eom_mean_const}) reduce to classical Tolman-Ehrenfest relations:
\begin{eqnarray*}
\label{TE_class_limit}
\left<E^{\mu}\right>=0,\qquad \left<B^{\mu}\right>=0,\qquad
\left<B^{\mu}E^{\nu}\right>=0,\\[1ex]
\left<E^{\mu} E^{\nu}\right>=-\frac{E^2}{3}h^{\mu\nu},\quad\mbox{and}\quad\left<B^{\mu}B^{\nu}\right>=-\frac{B^2}{3}h^{\mu\nu}.
\end{eqnarray*}
This calculation provides an explicit check that our generalized averaging reduces to the standard prescription when nonlinear coefficients vanish.

\section{The quasi-Magnetic Universe (qMU) model}
We shall choose a model to work with consisting of quadratic corrections in Maxwell's Lagrangian, which is the simplest case that can illustrate the modifications caused by the nonlinearities in a cosmological scenario and the consequences of the average process proposed in the previous section. This model is interesting because it corresponds to the first-order terms of well-known nonlinear theories such as Born-Infeld \cite{Born1934} and Euler-Heisenberg \cite{heisenberg, Herrera, chen}. The general case would be
\begin{equation}
\label{lagrangeana}
\mathcal{L}(F,G) = -\frac{1}{4}F + \alpha F^2 + \beta G^2 + \gamma FG,
\end{equation}
where $\alpha$, $\beta$, and $\gamma$ are the free parameters to be determined a posteriori. Models with $\beta,\gamma=0$ have been thoroughly studied in the literature, establishing a recipe to construct cyclic and bouncing cosmological models and inflationary phases (see \cite{Novello2009, Medeiros2012,Novello2007, Lorenci2002} and references therein). It is known that in standard QED, the numerical values of the Euler–Heisenberg coefficients are suppressed by inverse powers of the electron mass and are therefore extremely small on cosmological scales; consequently, the quadratic corrections in vacuum QED alone would not typically produce large early-universe backreaction. Our aim in this paper is primarily qualitative: to chart the dynamical possibilities that NLED terms can introduce. If one wishes to obtain quantitatively significant cosmological effects at early times, larger effective couplings or other high-energy mechanisms would be required (e.g., strong coupling sectors, additional fields, or effective models arising from beyond-QED physics). Moreover, a quantitative confrontation with observational bounds (CMB anisotropies, magnetogenesis constraints) is an important next step, and we now state this explicitly as future work.

Recalling that $F$ is only a function of time, we have $\left<F\right>=F$. However, $G$ may have an explicit random angular dependence, and thus we can compute the trace of the first compatibility equation (\ref{escolha1}), obtaining
\begin{equation}
\label{1st_ce_m_2}
\left(-\frac{1}{4}+2\alpha F\right)\left<G\right> + \gamma \left<G^2\right>=0.
\end{equation}
With this relation, the mean value of $G$ must be tied to its second moment. On the one hand, this is useful because all other averages can ultimately be expressed in terms of $\langle G\rangle$. On the other hand, there is no clear criterion for fixing $\langle G\rangle$ in a cosmological context without explicitly solving Maxwell’s equations, except in the special case $\gamma=0$, which forces $\langle G\rangle=0$. For this reason, we assume that the mixed term $FG$ is absent, and that $\langle G^2\rangle$ can instead be determined by independent arguments, as discussed below. Under this assumption, the tensor and vector compatibility equations reduce to the standard Tolman relations, whereas the scalar equations acquire additional contributions from the averaging procedure. In general, such extra terms always arise when the Lagrangian can be written in the separable form $\mathcal{L}(F,G)=\mathcal{L}_1(F)+\mathcal{L}_2(G)$. The remaining equations then read
\begin{eqnarray}
&&\left<\rho\right> = \frac{1}{4}F - \alpha F^2 + \beta\left<G^2\right> + E^2\left(1 - 8\alpha F \right),\label{rho_mean_model}\\[1ex]
&&\left<p\right> = -\frac{1}{4}F + \alpha F^2 - \beta\left<G^2\right> + \frac{1}{3}(F+E^2)\left(1 - 8\alpha F \right).\label{p_mean_model}
\end{eqnarray}
Using that now $\left<E^{\mu}\right>=0=\left<B^{\mu}\right>$ and also $\left<E^{\mu}B^{\nu}\right>=0$, we can make a further hypothesis saying that the electric and magnetic fields can be treated as independent random variables, and calculate 
\begin{equation}
\fl\qquad\langle G^2 \rangle = 16 \langle E^{\mu} B^{\nu} E^{\alpha} B^{\beta} \rangle g_{\mu\nu} g_{\alpha\beta}= 16 \langle E^{\mu} E^{\alpha}\rangle \langle B^{\nu} B^{\beta} \rangle g_{\mu\nu} g_{\alpha\beta}=\frac{16}{3} E^2B^2,
\end{equation}
where we used $\langle B^{\nu} B^{\beta} \rangle=-\frac{1}{3}B^2 h^{\nu\beta}$ and $\langle E^{\mu} E^{\alpha}\rangle=-\frac{1}{3} E^2 h^{\mu\alpha}$. Under these conditions, the equations of motion are identically satisfied on average in the cosmological cell. Nonetheless, it is still necessary to constrain the magnitude of the electric and magnetic fields, since we now have fewer equations than degrees of freedom, which leads us to the introduction of a qMU.

Cosmological observations and theoretical considerations constrain the limits on the magnitudes of primordial electric and magnetic fields. Current cosmological data, such as the CMB anisotropies and large-scale structure formation, place limits on the strength of primordial magnetic fields. Electric fields are generally not considered in the same way as magnetic fields because the early universe is expected to be highly conductive, which would quickly dissipate any large-scale electric field \cite{Planck2016,Durrer2013,Subramanian2016}, although there is a window in the pre-reheating phase where large electric fields are allowed, as we shall explain below.

This suggests that if both are present and not independent, a possible relation between their magnitudes is given by  
\begin{equation}  
\label{quasi_mag}  
E^2 = \sigma B^2,  
\end{equation}  
where $0 \leq \sigma \leq 1 $ from physical arguments. In equation (\ref{quasi_mag}) the symbols $E$ and $B$ denote the \emph{averaged}
magnitudes of the electric and magnetic fields measured in the comoving frame after applying the spatial averaging procedure described in section 2. In other words, $E^2\equiv\langle E_\mu E^\mu\rangle$ and $B^2\equiv\langle B_\mu B^\mu\rangle$ are scalar functions of time only. The effective, epoch-dependent, dimensionless parameter $\sigma$ therefore represents the electric-to-magnetic energy ratio on the cosmological cell. In the limit $\sigma \to 0 $, the results of \textit{Magnetic Universe} (MU) models are recovered \cite{Novello2009, Medeiros2012}, whereas the opposite limit, $ \sigma \to 1 $, corresponds to a purely radiative regime in which both EM invariants vanish on average. Henceforth, we refer to this scenario as the \textit{Statistical Null Universe} (SNU). It is also worth mentioning that during inflation or in the pre-reheating era the cosmic plasma is not necessarily a high-conductivity medium: conformal invariance of the electromagnetic sector can be broken \cite{TurnerWidrow1988,Ratra1992}, allowing the generation and survival of large-scale electric fields. Consequently, an averaged electric component need not be vanishingly small in those early epochs, and values of $\sigma$ that are non-negligible, in our model, can be physically motivated as characterizing the pre-reheating phase. After reheating the conductivity typically rises and large-scale electric fields are rapidly suppressed by charge screening and Ohmic dissipation, driving $\sigma$ toward small values at later times; our parametrization therefore captures both the (possibly nonzero) early-time electric contribution and its decay after reheating. In the literature, this phase is also referred to as inflationary magnetogenesis \cite{BambaYokoyama2004,MartinYokoyama2008,Kobayashi2014}.

Thus, for the qMU model, the expressions for the energy density and the isotropic pressure given by Eqs.\ (\ref{rho_mean_model}) and (\ref{p_mean_model}) reduce to
\begin{eqnarray}
&&\left<\rho\right> = \frac{1}{2}(1+\sigma)B^2 + \left[\tilde\beta - \tilde\alpha(1+3\sigma)\right]B^4,    \label{eq:rho_qm}\\[1ex]
&&\left<p\right>=\frac{1}{6}(1+\sigma)B^2 - \left[\tilde\beta +\frac{1}{3}\tilde\alpha(5-\sigma)\right]B^4,    \label{eq:p_qm}
\end{eqnarray}
where we introduce the auxiliary parameters $\tilde\alpha=4\alpha(1-\sigma)$ and $\tilde\beta=16\beta\sigma/3$. The null energy condition (NEC) and strong energy condition (SEC) can be readily computed, giving
\begin{eqnarray}
\left<\rho\right>+\left<p\right> = \frac{2}{3}\left[1-4\tilde\alpha B^2\right] \left(1+\sigma\right)B^2 \geq 0,\qquad \mbox{for NEC},\label{NEC}\\[1ex]
\left<\rho\right>+3\left<p\right> = \left(1+\sigma\right)B^2 -2\left[\tilde\beta+\tilde\alpha(3+\sigma)\right]B^4\geq0, \qquad \mbox{for SEC}.\label{SEC}
\end{eqnarray}
The NEC violation could occur if the magnetic field reaches values such that $B^2>\frac{1}{4 \tilde\alpha}$ for $\sigma\ne1$. However, the NEC can never be violated in the SNU. For the SEC, it is important to note that it can be violated for any choice of $\sigma$. This violation is necessary for achieving an accelerated expansion phase of the scale factor. The SEC is always violated once the magnetic field reaches the threshold $B^2 \geq \frac{1+\sigma}{2[\tilde\beta+\tilde\alpha(3+\sigma)]}$. However, for small values of $B^2$, the first term in equation (\ref{SEC}) dominates and the SEC is valid. 


The relation between the scale factor and the magnitude of the magnetic field can be obtained from the continuity equation
\begin{equation}
    \label{continuidade}
    \left<\dot{\rho}\right>+3\frac{\dot{a}}{a}\left(\left<\rho\right>+\left<p\right>\right)=0.
\end{equation}
By substituting equations (\ref{eq:rho_qm}) and (\ref{NEC}) into equation (\ref{continuidade}), we get after some manipulation
\begin{equation}
    \label{fatorequasimag}
    \frac{a}{a_0} = \sqrt{\frac{B_0}{B}}\left(\frac{1-4\tilde\alpha B^2}{1-4\tilde\alpha B_0^2}\right)^{w},
\end{equation}
where $w=(\tilde\beta-2\tilde\alpha\sigma)[4\tilde\alpha(1+\sigma)]^{-1}$ and we set $a(B=B_0)=a_0$. There are some interesting limits taken from (\ref{fatorequasimag}): when $\sigma=0$, it is possible to recover the results of the MU \cite{Novello2009} with
\begin{equation}
\label{fatoremag}
    \frac{a}{a_0} = \sqrt{\frac{B_0}{B}}.
\end{equation}
The expression (\ref{fatoremag}) is valid for any $\mathcal{L}(F)$ model, with vanishing electric field and the usual average procedure \cite{benaoum}; another interesting limit is $\sigma\rightarrow 1$. By solving equation (\ref{continuidade}) separately for this case or taking the appropriate limit in equation (\ref{fatorequasimag}), the scale factor reduces to
\begin{equation}
    \label{fatorenulo}
    \frac{a}{a_0}=\sqrt{\frac{B_0}{B}}e^{-\frac{8}{3}\beta\left(B^2-B_0^2\right)}.
\end{equation}
If the very early universe epoch allows for the presence of intense primordial electromagnetic fields, this behavior suggests that a rapid decrease in these fields would lead to an exponential growth of the scale factor. With this in mind, it would be interesting to confront this kind of model with cosmological data to investigate its viability in explaining inflation mechanics. However, it shall be left for a more observational work in the future.

\section{The dynamical system}

When the conditions in Eqs.\ (\ref{escolha1}) and (\ref{escolha2}) hold, the background dynamics is driven by the Friedmann equation
\begin{equation}
\label{friedmann1}
\left(\frac{\dot{a}}{a}\right)^2+\frac{\epsilon}{a^2}=\frac{\left<\rho\right>}{3},
\end{equation}
where $\epsilon$ is a constant positive, negative, or zero, depending on the spatial curvature (closed, open, or flat, respectively), and the acceleration equation
\begin{equation}
\label{friedmann2}
\frac{\ddot{a}}{a}=-\frac{1}{6}\left(\left<\rho\right>+3\left<p\right>\right).
\end{equation}
The latter determines whether the scale factor undergoes an accelerated growth and, together with the continuity equation (\ref{continuidade}), recovers equation (\ref{friedmann1}). Notably, acceleration occurs ($\ddot a>0$) only if the SEC (\ref{SEC}) is violated---a feature that will be shown to emerge in the models under consideration.

The solutions to Eqs.~(\ref{friedmann1}) and (\ref{friedmann2}) fully describe the evolution of the scale factor. However, the explicit time dependence of the energy density (\ref{eq:rho_qm}) and isotropic pressure (\ref{eq:p_qm}) remains undetermined, due to the lack of a known analytical solution for the magnetic field function $B(t)$. Consequently, direct integration of the Friedmann equation to find $a(t)$ is not feasible in general.

To study the evolution of both the magnetic field and the scale factor without solving the Friedmann equation explicitly, we construct a planar autonomous dynamical system based on equation (\ref{friedmann2}). This allows for a qualitative analysis of the magnetic field and, where possible, of the scale factor as well. 

By rewriting the left-hand side of equation~(\ref{friedmann2}) in terms of $B$ and its time derivatives—using relation (\ref{fatorequasimag}), and introducing the new variable $y = \dot{B}$, we obtain the dynamical system
\begin{equation}
\label{sistemaquasimag}
    \dot{B} = y,\qquad    \dot{y} = \frac{f_2(B) y^2 + f_3(B)}{f_1(B)},  
\end{equation}
where the auxiliary functions are given by
\begin{eqnarray}
f_1(B)&=&\frac{1}{2}B(1-4\tilde\alpha B^2)[1-4(1-4w)\tilde\alpha B^2],\label{f1}\\[1ex]
f_2(B)&=&\frac{3}{4}(1-4\tilde\alpha B^2)^2+64w(w-1)\tilde\alpha^2 B^4,\label{f2}\\[1ex]
f_3(B)&=&\frac{(1+\sigma)(1-4\tilde\alpha B^2)^2B^4\left[1 -2\tilde\alpha(4w+3)B^2\right]}{6}.\label{f3}
\end{eqnarray}

The phase space is constrained to the semi-plane $(B \geq 0, y)$. The function $f_1(B)$ may vanish at certain points, leading to a piecewise continuous vector field. As $f_1(B) \to 0$, the vector field $(\dot{B}, \dot{y})$ becomes effectively vertical, indicating singular behavior along the lines defined by
\begin{equation}
\label{eq:lines_div_f1}
B_1=0,\quad B_2=\frac{1}{2\sqrt{(1-4w)\tilde\alpha}}, \quad \mbox{or}\quad B_3= \frac{1}{2\sqrt{\tilde\alpha}}.
\end{equation}
The component $\dot{y}$ may change sign depending on the numerator, affecting the system’s behavior near these critical lines. This can be better understood by performing a time reparametrization $t \to \tau = \int dt / f_1(B)$, yielding a regularized system
\begin{equation}
\label{reg_sistemaquasimag}
    \frac{dB}{d\tau} = y f_1 (B), \quad \mbox{and}\quad \frac{dy}{d\tau} = f_2(B) y^2 + f_3(B).
\end{equation}
In this form, equilibrium points may exist on the lines where $f_1(B) = 0$, provided $y = \pm \sqrt{-f_3/f_2}$. Away from the singularities, these curves act as separatrices in both the regularized and original systems, with trajectories asymptotically approaching the singular lines as $\tau \to \infty$. As such, they should be interpreted primarily as parametrization singularities of the dynamical system; observable quantities as $\left<\rho\right>$ and $\left<p\right>$, given by equations (\ref{eq:rho_qm}) and (\ref{eq:p_qm}), respectively, remain finite as long as the magnetic field does not diverge.

In the analysis that follows, we examine the vector field near equilibrium points of the original system, investigate the flow behavior close to the divergence lines, and explore the sensitivity of solutions to the parameters $\tilde\alpha$, $w$ (or equivalently $\tilde\beta$), and $\sigma$. Finally, we constrain the phase portrait to those trajectories that satisfy the Friedmann equation. This approach also lays the groundwork for future studies on the system’s response to small perturbations.

\subsection{Equilibrium points and stability}

The equilibrium points of the dynamical system (\ref{sistemaquasimag}) are determined by setting $y=0$ and solving $f_3(B)=0$. These fixed points lie along the $B$-axis  and, depending on the value of $w$, there can be up to three real solutions
\begin{equation}
    \label{equilibrioquasimag}
    B_{P_1} = 0, \quad \mbox{and} \quad B_{P_2} = \frac{1}{\sqrt{2\tilde{\alpha}(4 w+3)}}, \quad \mbox{and} \quad B_{P_3} = \frac{1}{2\sqrt{\tilde{\alpha}}}.
\end{equation}
Note that the points $(B_{P_1},0)$ and $(B_{P_3},0)$ lie on the vertical divergence lines, which are defined by the roots of $f_1$, and this holds independently of the value of $w$. The point $B_{P_2}$, however, is not associated with the divergence of $f_1$; it corresponds solely to a root of $f_3$.

Depending on the parameter $y$ (with $\tilde\alpha>0$ fixed), the hierarchy of the critical values $B_i$ and the associated qualitative behavior of the system vary significantly. Table \ref{tab:poss_quali} summarizes the ordering of the divergence lines and equilibrium points, along with the type of stability associated with each equilibrium.

\begin{table}[ht]
    \centering
    \begin{tabular}{|c|c|c|c|c|}
    \hline
    Case & $w$ value & hierarchy & $B_{P_2}$ & $B_{P_3}$\\
    \hline
         I& $w<-\fracc{3}{4}$ & \quad $B_{1}<B_{2}<B_{3}$ & $\nexists$ & saddle\\
         II& $-\fracc{3}{4}<w<-\fracc{1}{4}$ & $B_{1}<B_{2}<B_{3}<B_{P_2}$ & center & saddle\\
         III& $-\fracc{1}{4}<w<-\fracc{1}{12}$ & $B_1<B_{2}<B_{P_2}<B_{3}$ & saddle & center\\
         IV& $-\fracc{1}{12}<w<0$ & $B_1<B_{P_2}<B_{2}<B_{3}$ & center & center\\
         V& $0<w<\fracc{1}{4}$ & $B_1<B_{P_2}<B_{3}<B_{2}$ & center & saddle\\
         VI& $w>\fracc{1}{4}$ & $B_1<B_{P_2}<B_{3}$, \, $\nexists\, B_{2}$ & center & saddle\\ \hline
    \end{tabular}
    \caption{Relative position of divergence lines, equilibrium points, and their associated stabilities for different values of $w$ and $\tilde\alpha>0$.}
    \label{tab:poss_quali}
\end{table}

To understand the dynamics near the equilibrium points, we linearize the system (\ref{sistemaquasimag}) using the Jacobian matrix:
\begin{equation}
   \label{jacobiana}
    \mathcal{J}\big|_{\textrm{eq. pts.}} = \left|\begin{array}{cc}
0 & 1 \\
\frac{f'_3}{f_1}-\frac{f_3f'_1}{f_1^2} & 0
    \end{array}\right|,
\end{equation}
where we use the fact that all equilibrium points have $y=0$. Evaluated at each equilibrium point, this yields:
\begin{equation}
\label{jac_eq}
\fl\mathcal{J}\big|_{B_{P_1}} = \left|\begin{array}{cc}
0 & 1 \\
0 & 0 
\end{array}\right|, \quad \mathcal{J}\big|_{B_{P_2}} = \left|\begin{array}{cc}
0 & 1 \\
\frac{-(1+4w)(1+\sigma)}{3\tilde\alpha (4w+3)(1+12w)} & 0 
\end{array}\right|, \quad \mathcal{J}\big|_{B_{P_3}} = \left|\begin{array}{cc}
0 & 1 \\
\frac{(1+4w)(1+\sigma)}{48w\tilde\alpha} & 0 
\end{array}\right|.
\end{equation}
The eigenvalues of each matrix are determined by the square roots of the lower-left entries. A negative value yields a center, and a positive value indicates a saddle point. The behavior of the equilibrium points is summarized in table \ref{tab:poss_quali}. Special attention must be given to the origin $(0,0)$, since its linearization yields a non-hyperbolic fixed point. Specifically, expanding $\dot y$ near $B=0$ with $y\neq0$ yields a singular expression: $\dot y\sim y^2/B$. This suggests the origin is not Lyapunov stable (see e.g., \cite{Perko2001,Wiggins2003}), and the vector field is not continuous at this point, which precludes the use of classical linearization theorems. Remarkably, the Friedmann equation with $\epsilon\leq0$  can be interpreted as a family of center manifolds. According to the Center Manifold Theorem \cite{kuznetsov2004}, the solutions in this case will approach the origin as an attractive node when $\dot B<0$, and move away from it as a repulsive node when $\dot B>0$. For $\epsilon>0$, the solutions instead exhibit the characteristic behavior of a local saddle.

To properly analyze the behavior near the divergence lines, particularly those that coincide with equilibrium points, we consider a time reparametrization introduced in equation (\ref{reg_sistemaquasimag}). This transformation makes the field continuous and permits a well-defined phase-space analysis near singularities of the original system.
As we said, this system admits additional equilibrium points off the $y=0$ axis given by:
\begin{equation}
    \label{newequipoints}
    P_\pm = \left(B_2,\pm\frac{1}{4\tilde{\alpha}(1-4w)^2}\sqrt{\frac{-w(1+\sigma)(1+12w)}{3}}\right),
\end{equation}
which exist only for $-1/12<w<0$. The linearization now gives
\begin{equation}
    \label{jacobianPpm}
    \mathcal{J}(P_{\pm}) = \left|\begin{array}{cc}
 \frac{4w}{1-4w} y_\pm & 0 \\
f'_2y_\pm^2+f'_3 & \frac{-8w}{1-4w} y_\pm 
    \end{array}\right|.
\end{equation}
The signs of the eigenvalues (diagonal entries) confirm the saddle nature of these points, with the direction of approach determined by the sign of $y_{\pm}$.

The case $\tilde\alpha<0$ eliminates one equilibrium point and one separatrix, making the qualitative analysis simpler and without introducing new features. For completeness, however, we present the detailed treatment in \ref{app:negative_alpha}.

\subsection{The constraint equation and the physical universes}

Using the relation (\ref{friedmann1}), we can derive a constraint equation for the dynamical system. Since equation (\ref{friedmann2}) is a second-order differential equation, it admits solutions that may not satisfy the first-order constraint given by (\ref{friedmann1}). By rewriting the left-hand side of (\ref{friedmann1}) with the help of (\ref{fatorequasimag}), we obtain the constraint equation:
\begin{equation}
     \label{equaçãodevinculo}
\fl    \frac{\left[1-4\tilde{\alpha}\left(1-4w\right)B^2\right]^2y^2}{4 B^2\left(1-4\tilde{\alpha} B^2\right)^2}+\frac{\epsilon B}{a_0^2 B_0} \left(\frac{1-4\tilde{\alpha}B_0^2}{1-4\tilde{\alpha}B^2}\right)^{2w}=\frac{(1+\sigma)B^2}{6}\left[1-2\tilde{\alpha}(1-4w)B^2\right].
\end{equation}
This equation shows how the initial conditions $B_0$ and $a_0$, and the spatial curvature $\epsilon$ influence the evolution of the magnetic field. In particular, when $\epsilon\neq0$, the choice of $B_0$ sets the region where physical solutions occur, since the positivity of the term in parentheses is related to the separatrix $B_3$ for arbitrary $w$. Once $B_0$ is fixed, the sign of this term is determined by $\epsilon$. Furthermore, for $\epsilon\geq0$ and $w<1/4$, the solutions have an upper limit in the magnetic field given by $B=\sqrt{2}B_2$, coming from the last term of equation (\ref{equaçãodevinculo}). This bound reflects the need for a positive energy density in the Friedmann equation for flat and closed universes. For $w>1/4$, $B_2$ is absent, and the energy density is always positive. 

In terms of matter content, a simple analysis of the energy density and the pressure shows a close relation between the critical values of the dynamical system (\ref{sistemaquasimag}) and the energy conditions. In particular, $\rho$ is positive for $B<[2\tilde\alpha(1-4w)]^{-\frac{1}{2}}$ and $p$ is positive when $B<[2\tilde\alpha(5+12w)]^{-\frac{1}{2}}$. The NEC is always satisfied within the region $B\leq B_3$ (see Eqs. \ref{NEC} and \ref{eq:lines_div_f1}). Since it coincides with a line of discontinuity of the flow, the solutions in the phase space are divided into those that are restricted to fully satisfy the NEC and those that never satisfy this energy condition. In contrast, the SEC is valid when $B\leq B_{P_2}$. As it is an equilibrium point, there are solutions that partially satisfy the SEC along the evolution. Furthermore, as the SEC depends on the parameter $w$, when $w<-3/4$ the SEC is satisfied in the whole phase portrait; for $-3/4<w<-1/4$ the SEC is valid in a region bigger than the NEC; for $w=-1/4$ they both coincide, and for $w>-1/4$ the SEC is valid in a region of the phase space smaller than the NEC. The implementation of these constraints to the dynamical system given by equation (\ref{sistemaquasimag}) is illustrated in the phase diagrams below.

\begin{figure}[ht]
    \centering
    \includegraphics[width=0.8\linewidth]{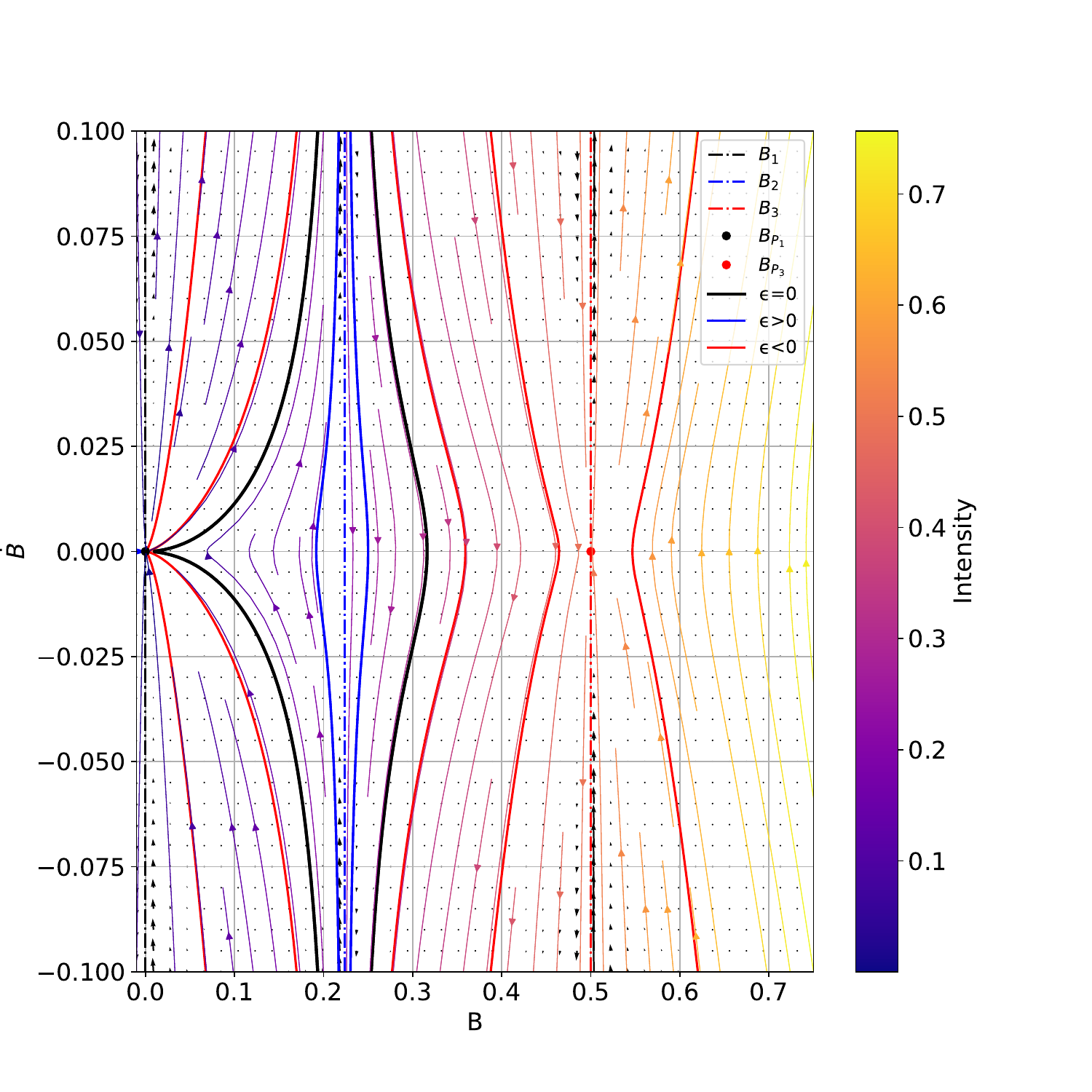}
    \caption{Phase Portrait of Case I. The vertical dot-dashed lines $B_1$ (black), $B_2$ (blue), and $B_3$ (red) are the separatrices. The equilibrium points are $B_{P_1}$ (node) and $B_{P_3}$ (saddle). $B_{P_2}$ is absent. The solid black line indicates the possible flat universes, separating the diagram into disjoint regions in terms of curvature. For this phase diagram, we choose $\tilde\alpha=1$, $\sigma=1/2$, without loss of generality, and $w=-1$.}
    \label{fig:w1n}
\end{figure}

In Case I, it is noticeable that the SEC is never violated. Additionally, the vertical line defined by $B_2$ separates two regions in which the solutions exhibit qualitatively distinct behaviors. For magnetic field values $B < B_2$, the field grows rapidly, approaching the asymptotic maximum at $B_2$. In this regime, the solutions tend to the equilibrium point at the origin (past or future) for flat and negative spatial curvature, while for positive spatial curvature, a minimum value of the magnetic field is evident. For $B > B_2$, all three curves exhibit similar behavior: they reach a maximum at a finite value $B<B_3$, without violating the NEC. Notably, the curve corresponding to negative spatial curvature ($\epsilon < 0$) reaches a maximum magnetic field value that exceeds the upper bound required for the positivity of the energy density.

\begin{figure}[ht]
    \centering
    \includegraphics[width=0.8\linewidth]{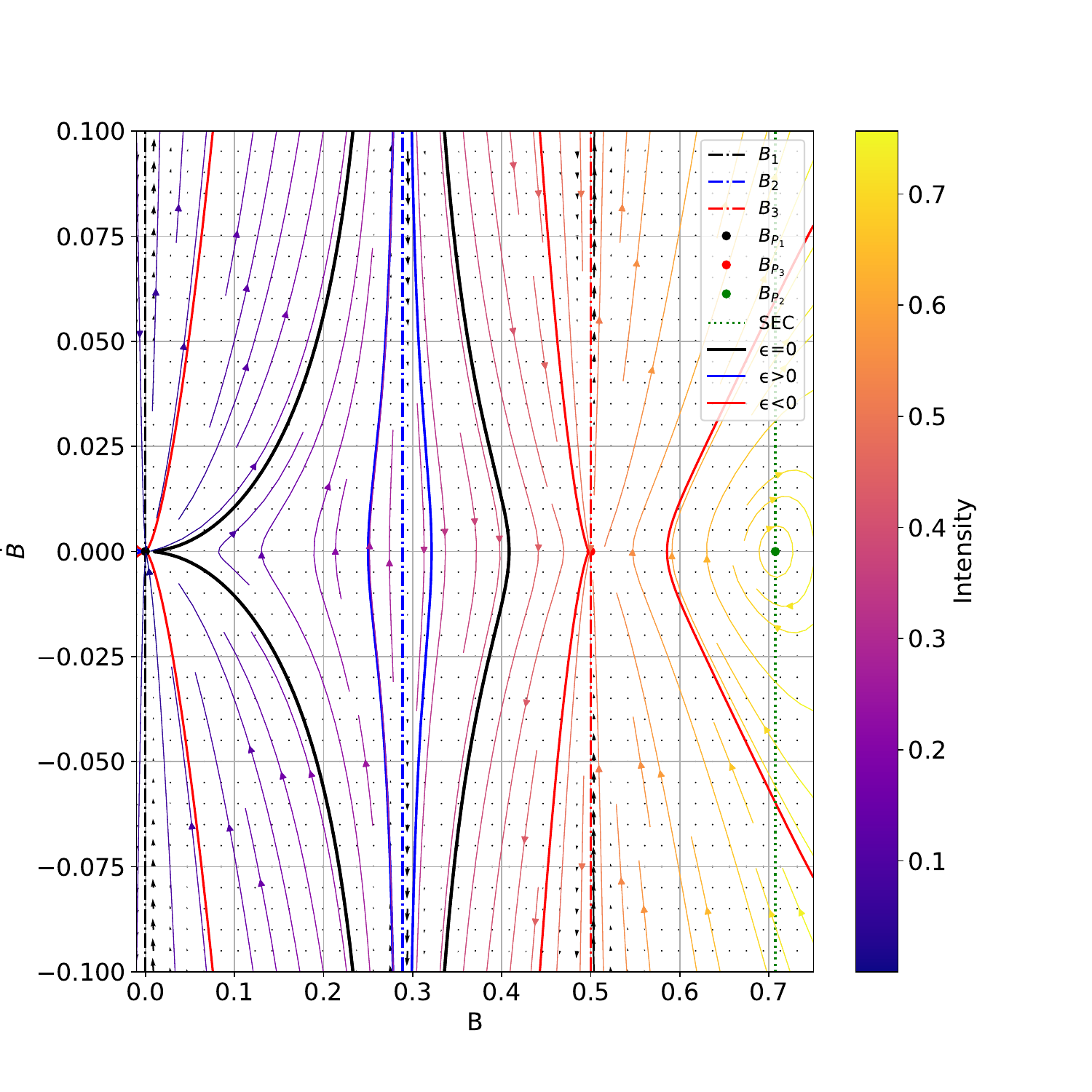}
    \caption{Phase Portrait of Case II. The vertical dot-dashed lines caption follows the previous case. The equilibrium points are now the $B_{P_1}$ (node), $B_{P_2}$ (center), and $B_{P_3}$ (saddle). Again, the flat universe solutions represented by the solid black line separate the diagram into disjoint regions in terms of curvature. For this phase diagram, we choose $\tilde\alpha=1$, $\sigma=1/2$, and $w=-1/2$.}
    \label{fig:w0.5n}
\end{figure}

Understanding the evolution of the magnetic field provides insights into the qualitative behavior of the scale factor, as described by equation (\ref{fatorequasimag}). When $w < 0$, maxima in the magnetic field correspond to minima in the scale factor, and vice versa. Furthermore, when $B = 0$ or $B = B_3$, the scale factor tends to grow indefinitely. If the magnetic field becomes arbitrarily large, the scale factor is expected to approach a singular value. However, in Case I (figure \ref{fig:w1n}), this situation never occurs for $B < B_3$. In contrast, for $\epsilon < 0$ and $B > B_3$, there is no upper bound for the magnetic field, suggesting that the scale factor may evolve toward a singularity.

\begin{figure}[ht]
    \centering
    \includegraphics[width=1\linewidth]{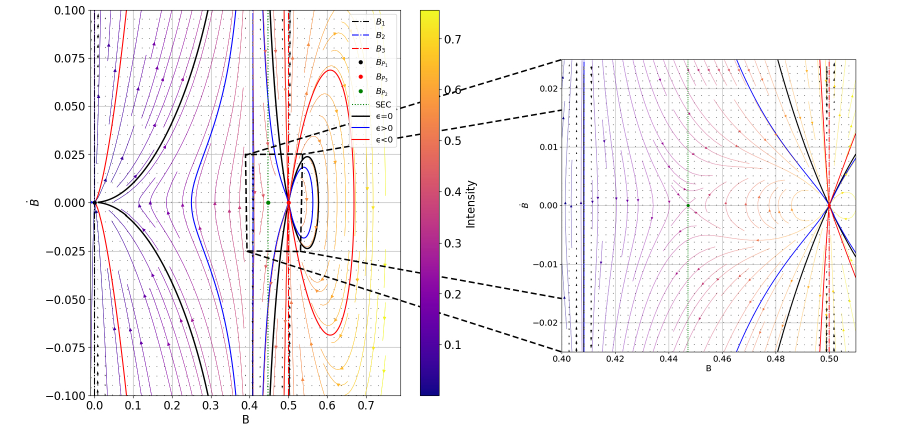}
    \caption{Phase Portrait of Case III. The vertical dot-dashed lines caption follows the previous case. The equilibrium points are the $B_{P_1}$ (node), $B_{P_2}$ (saddle (see the zoomed region on the left panel)), and $B_{P_3}$ (``center''). Again, the flat universe solutions represented by the solid black line separate the diagram into disjoint regions in terms of curvature, but now there is a branch on the right of $B_3$. For this phase diagram, we choose $\tilde\alpha=1$, $\sigma=1/2$, and $w=-1/8$.}
    \label{fig:w0.125n}
\end{figure}

Case II (figure \ref{fig:w0.5n}) exhibits behavior similar to Case I regarding the constraint curves. However, it now reveals the presence of the equilibrium point $B_{P_2}$, which marks the boundary for the validity of the SEC. Notably, the curves corresponding to $\epsilon < 0$ can reach  $B_{P_2}$, indicating that the scale factor may grow indefinitely—potentially signaling a ``Big Rip'' scenario.

As previously mentioned, $B_{P_2}$ is now visible in the phase diagram, which allows the trajectories with $B > B_3$ and $\epsilon < 0$ to form closed trajectories. This behavior can be interpreted as a cyclic universe, with both a minimum and a maximum scale factor occurring at finite values. It is also important to note that solutions in this region always violate the NEC and may also violate the SEC.

In Case III (figure \ref{fig:w0.125n}), the main change in behavior arises from the fact that the SEC can be violated before the NEC, as $B_{P_2} < B_3$. When $B_{P_3}$ lies between both vertical lines, it acts as a saddle point, as shown in the zoomed panel from figure \ref{fig:w0.125n}, enabling some curves with $\epsilon > 0$ to form a close orbit originating from $B_{P_3}$. Additionally, all curves independently of $\epsilon$ reach $B_{P_3}$ from the left-hand side, which indicates a divergence in the scale factor. For the solutions on the right-hand side of $B_3$, now any spatial curvature is possible. These curves each reach a maximum at a finite value of $B$, but they ultimately converge back toward $B_{P_3}$.

\begin{figure}
    \centering
    \includegraphics[width=1\linewidth]{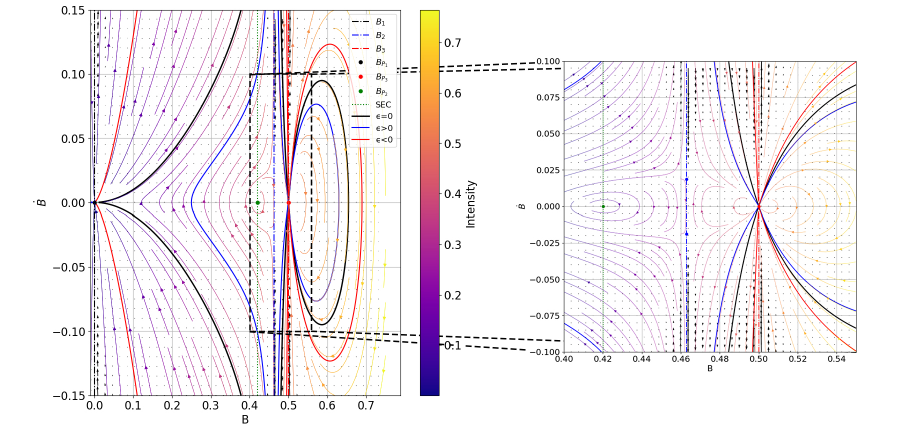}
    \caption{Phase Portrait of Case IV. The vertical dot-dashed lines caption follows the previous case. The equilibrium points are the $B_{P_1}$ (node), $B_{P_2}$ (center), and $B_{P_3}$ (``center''). The zoomed region also shows the appearance of two saddle points (blue) along $B_2$ whose separatrices delineate the attraction basin of $B_{P_3}$. Flat universe solutions separate the diagram again into disjoint regions, and the branch on the right of $B_3$ remains. For this phase diagram, we choose $\tilde\alpha=1$, $\sigma=1/2$, and $w=-1/24$.}
    \label{fig:w0.042n}
\end{figure}

In Case IV (figure \ref{fig:w0.042n}), the equilibrium point $B_{P_2}$ appears to the left of $B_2$. This configuration allows for the formation of a closed orbits region with $\epsilon > 0$ around $B_{P_2}$ whose boundary is established by the separatrices of two saddle points that emerge symmetrically located at $B_2$, as illustrated in the zoom of figure \ref{fig:w0.042n}. Between $B_2$ and $B_3$, the separatrices of the saddle point connect to $B_{P_3}$, creating a closed loop region originating in $B_{P_3}$. In the complementary regions, the behavior remains consistent with that of the previous cases, as these trajectories lie outside the influence of the attraction basin mentioned above. The dynamics for $B > B_3$ are analogous to those observed in Cases II and III.

\begin{figure}
    \centering
    \includegraphics[width=0.8\linewidth]{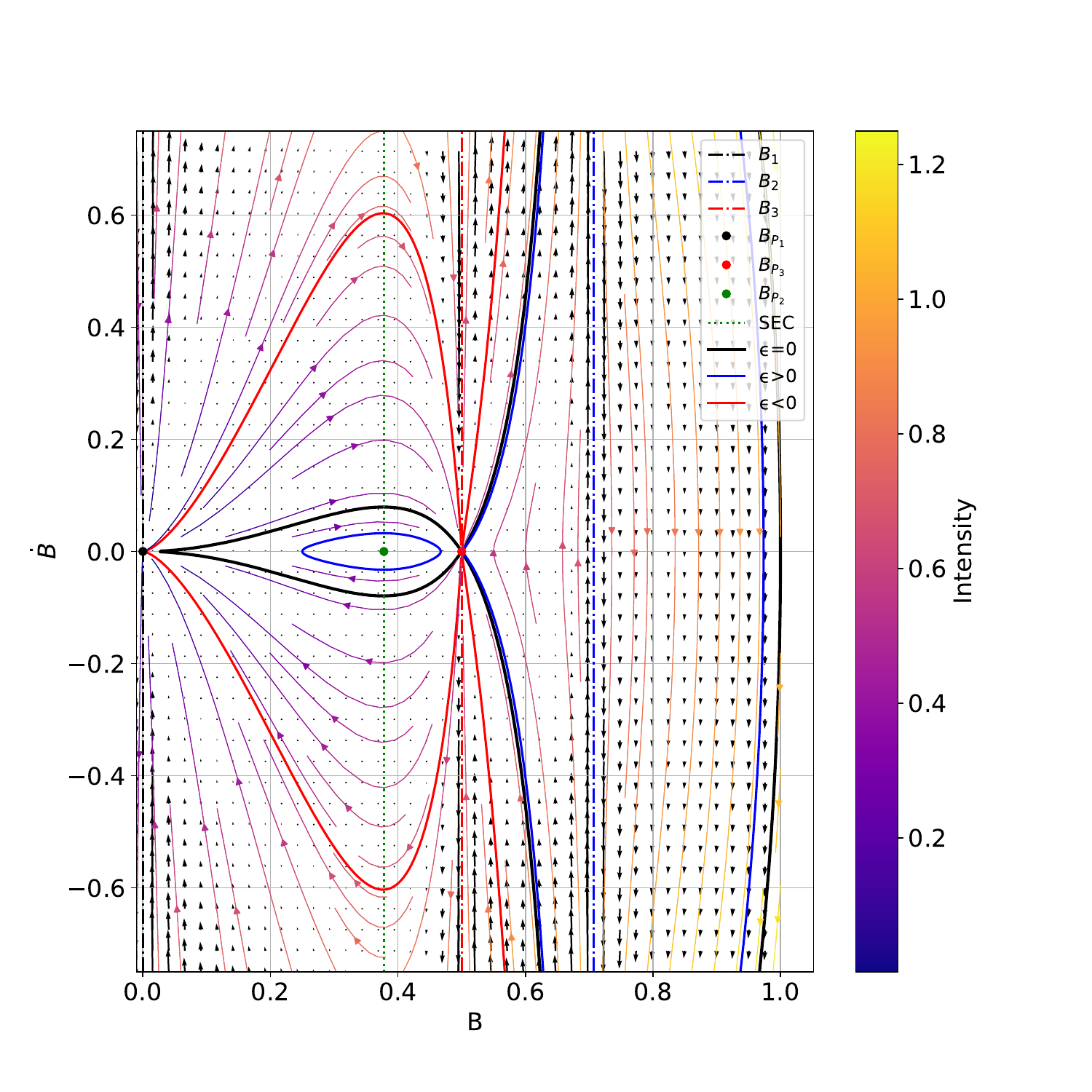}
    \caption{Phase Portrait of Case V. The vertical dot-dashed lines caption follows the previous case. The equilibrium points are $B_{P_1}$ (node), $B_{P_2}$ (center), and $B_{P_3}$ (saddle). Flat universe solutions continue separating the diagram into disjoint regions, but now closed orbits are impossible on the right side of $B_3$. For this phase diagram, we choose $\tilde\alpha=1$, $\sigma=1/2$, and $w=1/8$.}
    \label{fig:w0.125p}
\end{figure}

\begin{figure}
    \centering
    \includegraphics[width=0.8\linewidth]{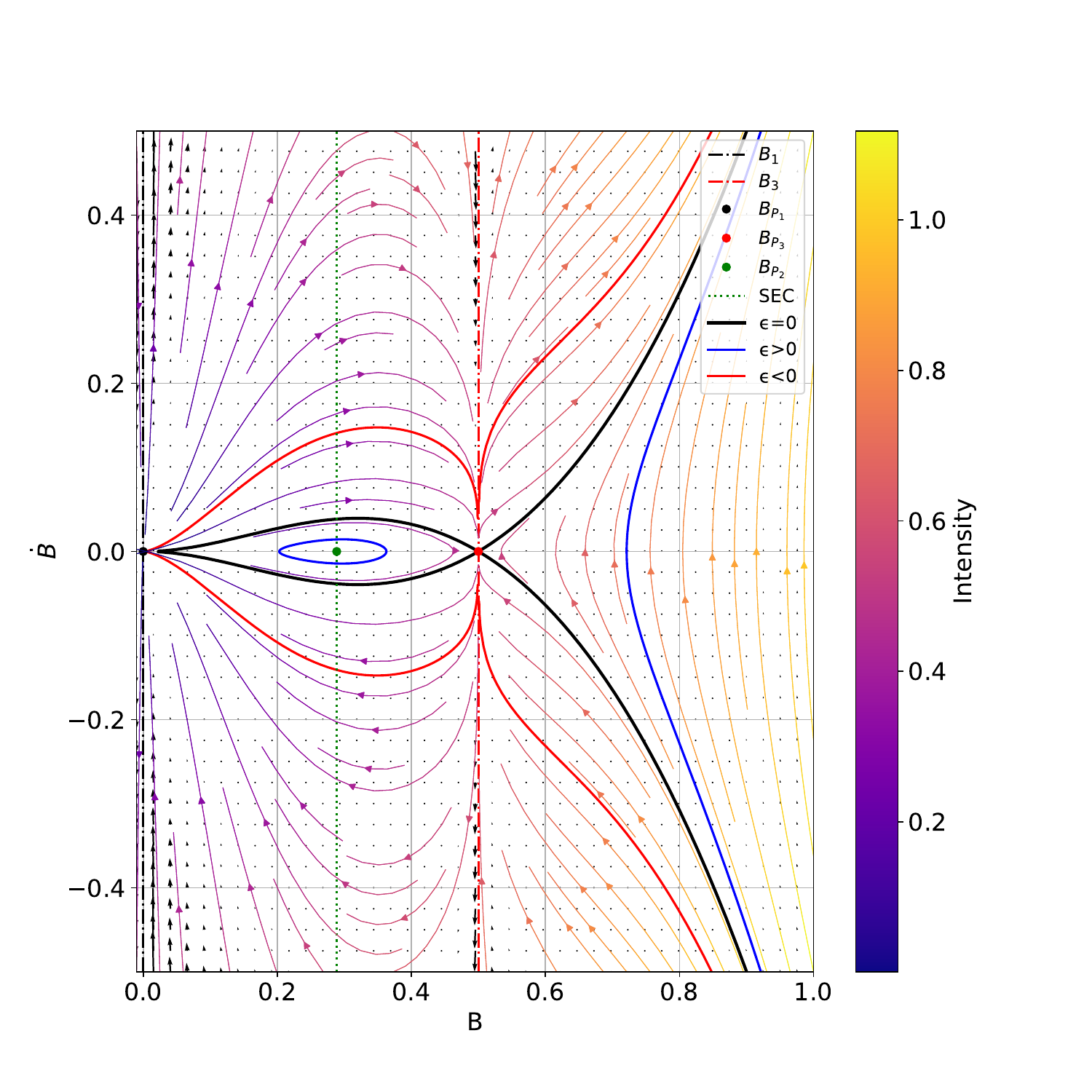}
    \caption{Phase Portrait of Case VI. The only difference from the previous case is that the region on the right of $B_2$ is suppressed since this line is pushed away to infinity. For this phase diagram, we choose $\tilde\alpha=1$, $\sigma=1/2$, and $w=3/4$.}
    \label{fig:w0.75p}
\end{figure}

For $w > 0$, two distinct situations arise: Case V (figure \ref{fig:w0.125p}) and Case VI (figure \ref{fig:w0.75p}). In Case V, the vertical line at $B_2$ now appears to the right of the phase diagram, acting as an asymptotic value toward which the curves converge and serving as a lower bound for trajectories where $B > B_2$. The curves with $B < B_3$ and $\epsilon \leq 0$ now reach a maximum at the equilibrium point $B_{P_3}$, indicating a divergent (singular) scale factor. In contrast, the curves with $\epsilon > 0$ form closed trajectories, suggesting cyclic behavior. In Case VI, the primary difference is the absence of the vertical line at $B_2$ (pushed away to infinity), which allows the magnetic field to grow indefinitely.

As mentioned above, the flat universe acts as a boundary in the phase portrait, separating regions according to the spatial curvature. This division, however, does not strongly restrict the scenarios allowed for nonzero curvature (see figures~\ref{fig:w0.125n}–\ref{fig:w0.125p}). For example, both cyclic and bouncing solutions can occur for $\epsilon>0$ or $\epsilon<0$ with the same set of parameters. Thus, although current cosmological observations favor a nearly flat universe \cite{Planck2020}, this should not be regarded as a severe limitation on the possible dynamics of the early universe.

\subsection{Special cases: the MU and SNU models}
\begin{figure}
\centering
\includegraphics[width=0.49\linewidth]{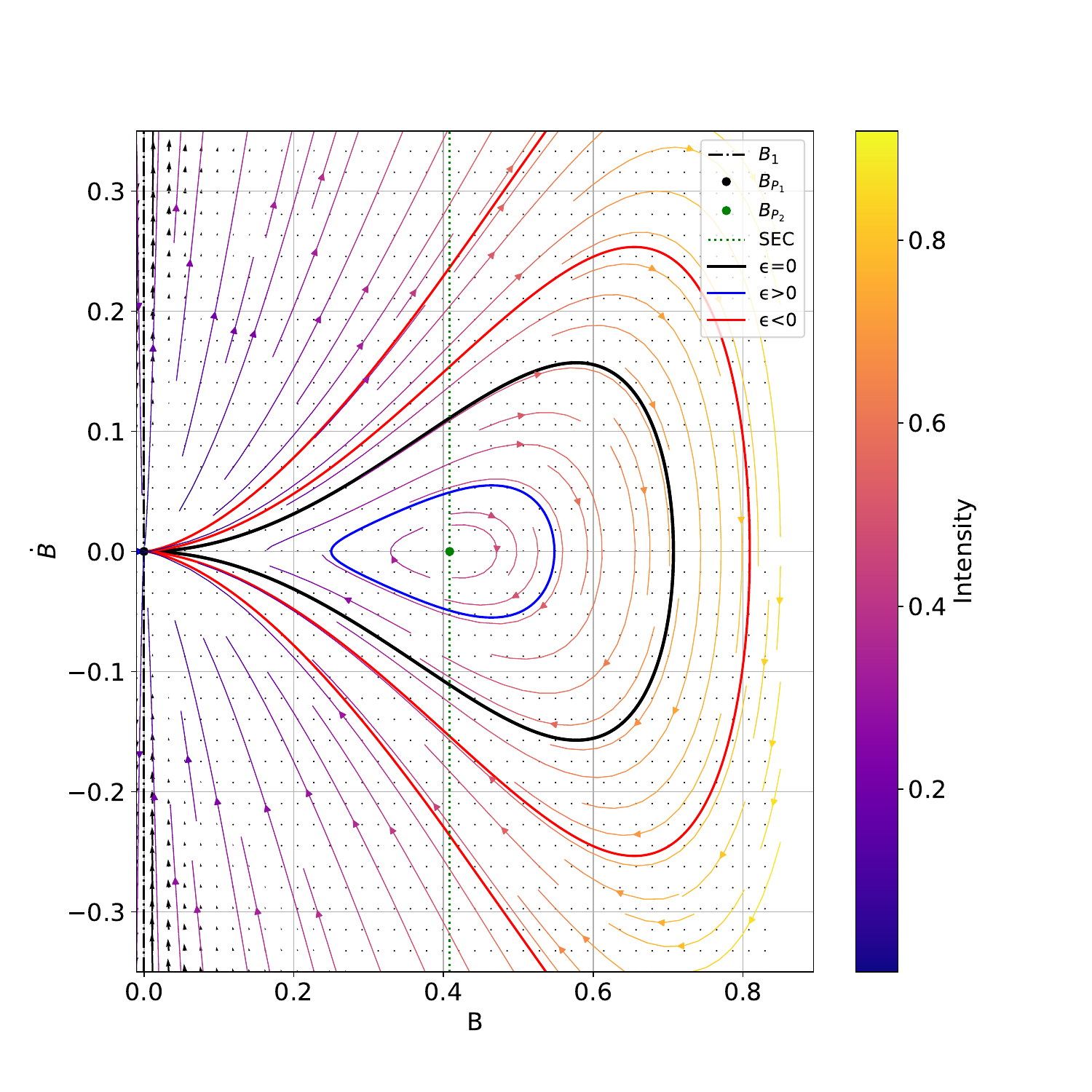}
\includegraphics[width=0.49\linewidth]{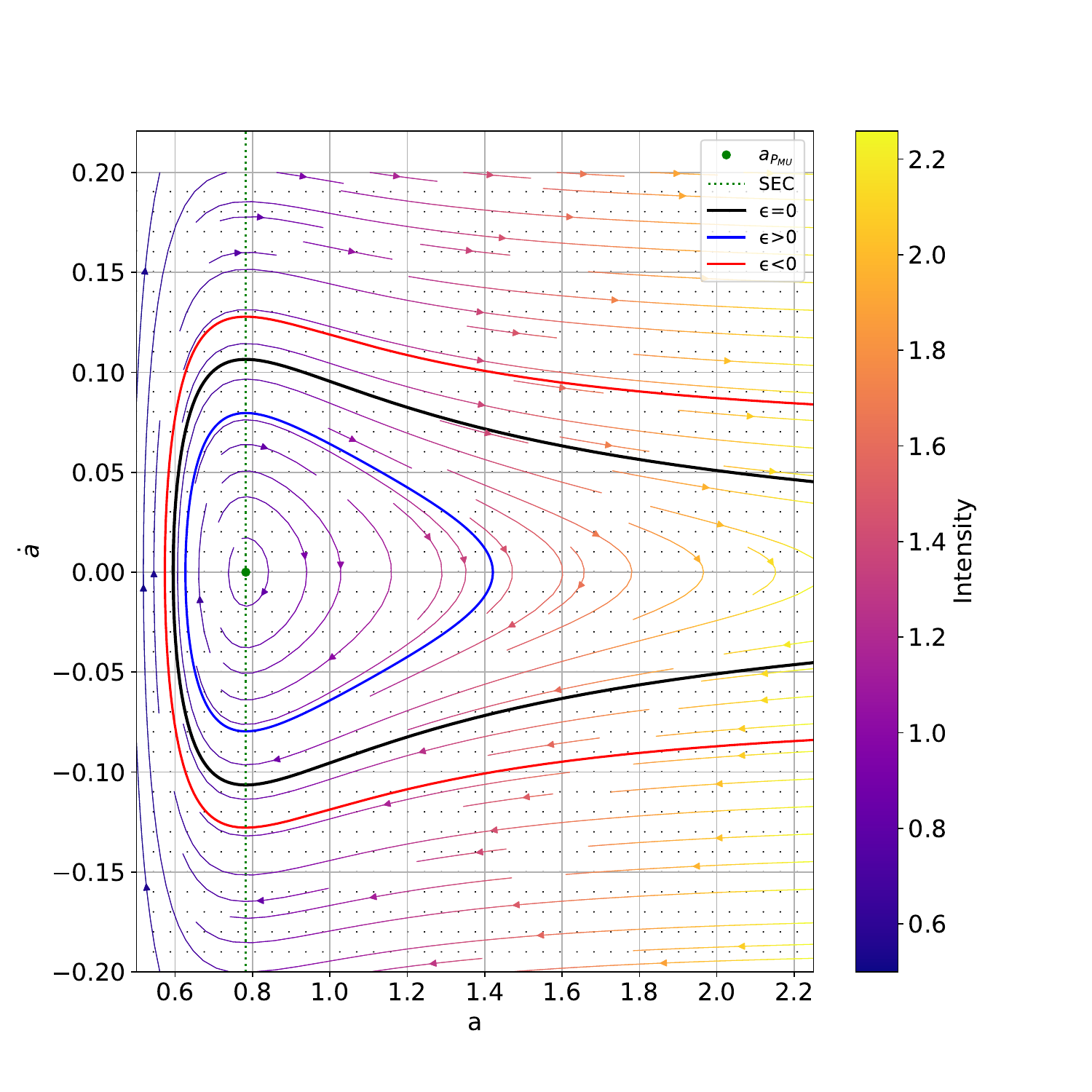}
\caption{Phase portraits for the MU case. Left: $(B, \dot{B})$ plane, showing equilibrium points $B_{P_1}$ and $B_{P_2}$ (both centers). Right: $(a, \dot{a})$ plane, with the equilibrium point $P_{MU}$ (center). Again, the solid black line indicates the possible flat universe solution, which separates the diagram into disjoint regions in terms of curvature. Here, we use $\tilde{\alpha} = 1$.}
\label{fig:w0n}
\end{figure}

In the MU case, we have $\sigma = 0$ (no electric field), which implies that $w=0$. With this assumption, the dynamical system given by equation (\ref{sistemaquasimag}) simplifies to
\begin{equation}
\label{sistemamag}
\dot{B} = y, \qquad \dot{y} = \frac{B^3}{3}\left(1 - 6\tilde{\alpha} B^2\right) + \frac{3y^2}{2B}.
\end{equation}
Apart from the origin, this system has another equilibrium point given by $B_{P_2}=1/\sqrt{6\tilde\alpha}$, whose linearization indicates it as a center (see the left panel of figure \ref{fig:w0n}).

In this case, the relation (\ref{fatoremag}) becomes invertible, allowing us to express the system in terms of the scale factor:
\begin{equation}
\label{sistemafatore}
\dot{a} = z, \qquad \dot{z} = -\frac{a_0^4 B_0^2}{6a^3} \left(1 - \frac{6\tilde{\alpha} a_0^4 B_0^2}{a^4}\right),
\end{equation}
and the corresponding constraint equation takes the form
\begin{equation}
\label{vinculofatore}
\left(\frac{z}{a}\right)^2 + \frac{\epsilon}{a^2} = -\frac{1}{6} \frac{a_0^4 B_0^2}{a^4} \left(1 - 2\tilde{\alpha} \frac{a_0^4 B_0^2}{a^4}\right).
\end{equation}

The equilibrium point in this formulation is given by
\begin{equation}
\label{equilibriofatoremag}
a_{P_{MU}} = a_0 (2\tilde{\alpha B_0^2})^{\frac{1}{4}}.
\end{equation}
The linearization of the system (\ref{sistemafatore}) around $a_{P_{MU}}$ yields a Jacobian matrix with eigenvalues $r_{\pm} = i/(6\sqrt{\tilde\alpha})$, indicating the presence of a center for all $\tilde\alpha > 0$.

The qualitative behavior indicates that the magnetic field reaches a maximum in finite time for all spatial curvatures. For the scale factor, a maximum occurs only when $\epsilon > 0$, while for $\epsilon \leq 0$, the universe expands indefinitely (see the right panel of figure \ref{fig:w0n}). Specifically, for $\epsilon < 0$, the scale factor grows at a constant rate asymptotically, whereas for $\epsilon = 0$, the growth rate approaches zero. In all three curvature scenarios, the scale factor exhibits a minimum, indicating a cosmological bounce. For the magnetic field, the presence of a minimum when $\epsilon > 0$ corresponds to the maximum in the scale factor, which in turn represents a re-bounce, leading to cyclic universes.

In the SNU case, the relation between the scale factor and the magnetic field is given by equation (\ref{fatorenulo}). Taking the limit $\sigma \rightarrow 1$, the dynamical system (\ref{sistemaquasimag}) reduces to
\begin{equation}
    \label{sistemanulo}
\fl\qquad    \dot{B} = y, \qquad \dot{y} = \frac{2 B^3}{3}\frac{(1-\tilde{\beta} B^2)}{(1+2\tilde{\beta}B^2)}+\frac{y^2 \left[(1+2\tilde{\beta}B^2)(3+2\tilde{\beta}B^2)-8\tilde{\beta}B^2\right]}{2B (1+2\tilde{\beta}B^2)}.
\end{equation}
The equilibrium points of this system are given by
\begin{equation}
    \label{equilibrionulo}
    B_{P_1} = 0, \quad \mbox{and} \quad B_{P_2} = \frac{1}{\sqrt{\tilde{\beta}}}.
\end{equation}
It should be mentioned that the presence of $\tilde\beta$ indicates the relevant contribution of $\left<G^2\right>$ even if $\left<G\right>=0$, once the correct average procedure is taken into account.

The relation (\ref{fatorenulo}) is also invertible, allowing the magnetic field to be expressed as a function of the scale factor:
\begin{equation}
    \label{campomagnulo}
    B (a) = \sqrt{\frac{\mathrm{W}\left[\frac{2 \tilde{\beta}B_0^2a_0^4  e^{2\tilde{\beta}B_0^2}}{a^4} \right]}{2\tilde{\beta}}}.
\end{equation}
where $\mathrm{W}$ denotes the Lambert $W$ function. 

This expression makes it possible to rewrite the system in terms of the scale factor:
\begin{equation}
    \label{sistemafatornulo}
    \dot{a} = z, \qquad \dot{z} = -\frac{a}{6\tilde{\beta}}\mathrm{W}\left[\frac{2 \tilde{\beta}B_0^2a_0^4  e^{2\tilde{\beta}B_0^2}}{a^4} \right]\left(1-\frac{\mathrm{W}\left[\frac{2 \tilde{\beta}B_0^2a_0^4  e^{2\tilde{\beta}B_0^2}}{a^4} \right]}{2}\right),
\end{equation}
subject to the constraint:
\begin{equation}
    \label{vinculonulo}
    \frac{z^2}{a^2}+\frac{\epsilon}{a^2} =\frac{\mathrm{W}\left[\frac{2 \tilde{\beta}B_0^2a_0^4  e^{2\tilde{\beta}B_0^2}}{a^4} \right]}{6\tilde{\beta}}\left(1+\frac{\mathrm{W}\left[\frac{2 \tilde{\beta}B_0^2a_0^4  e^{2\tilde{\beta}B_0^2}}{a^4} \right]}{2}\right).
\end{equation}

The corresponding equilibrium point in this formulation is
\begin{equation}
    \label{equilibrofatorenulo}
    a_{P_{SNU}} = a_0 \sqrt{B_0}\tilde{\beta}^{1/4} e^{\frac{\tilde{\beta}B_0^2-1}{2}}.
\end{equation}

\begin{figure}
    \centering
    \includegraphics[width=0.49\linewidth]{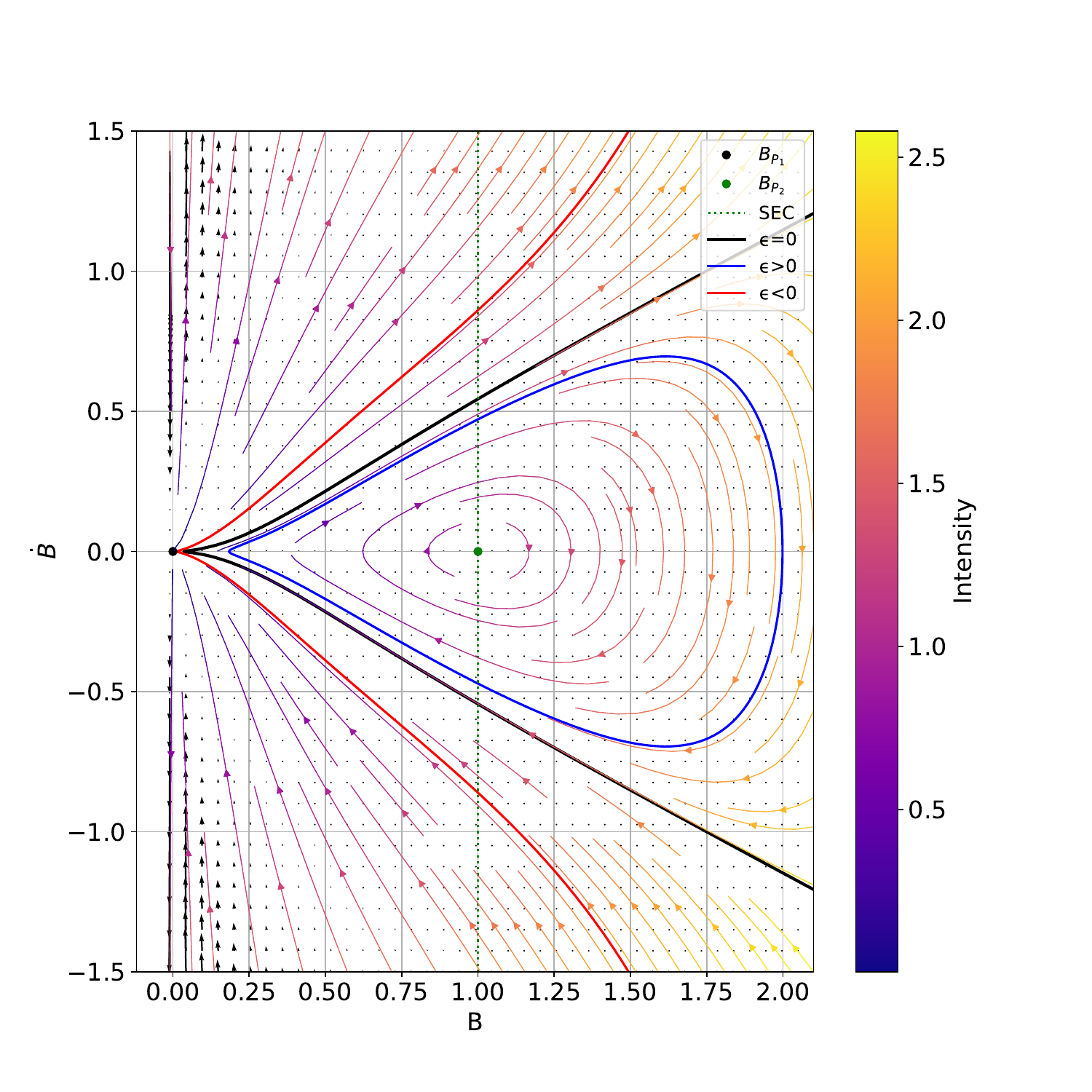}
    \includegraphics[width=0.49\linewidth]{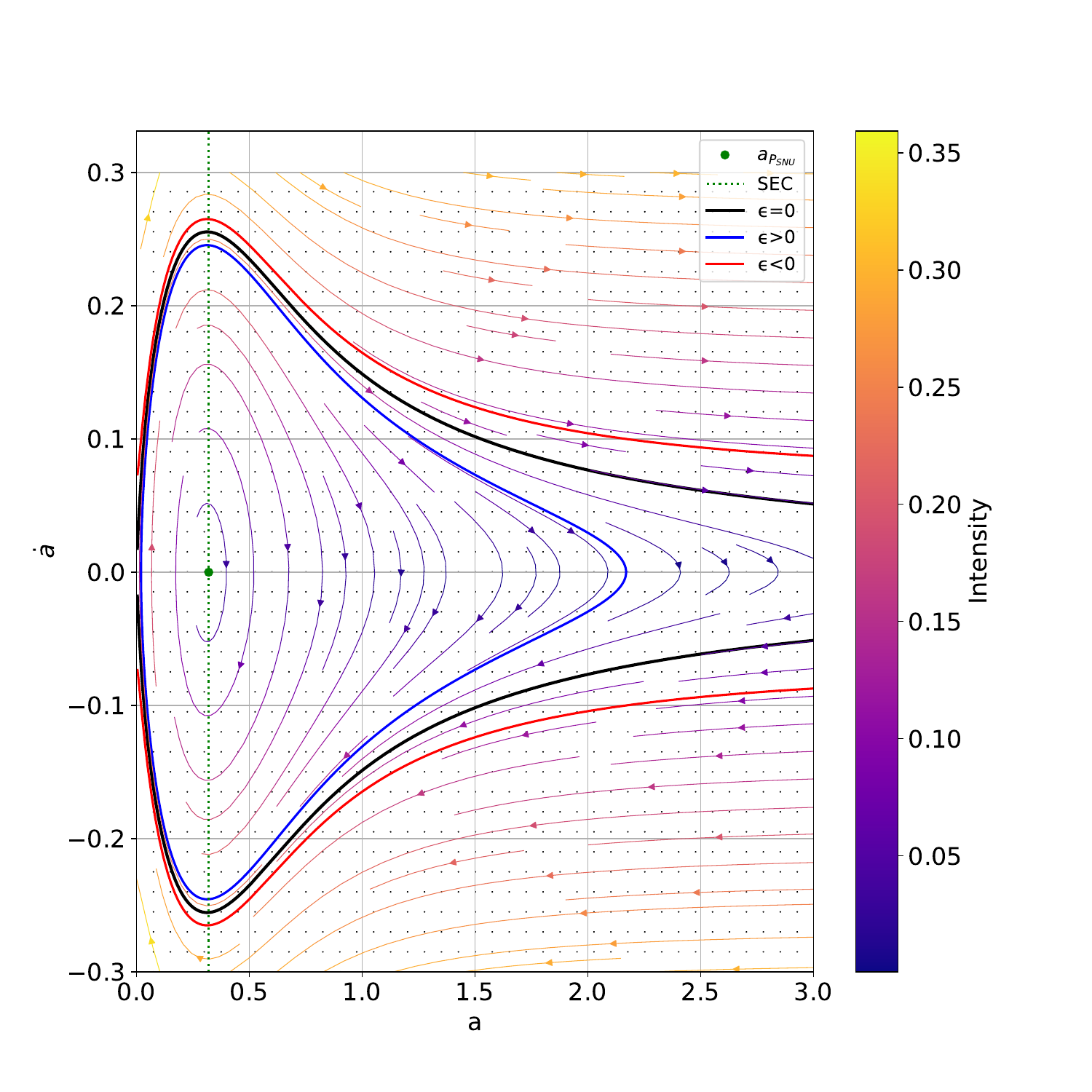}
    \caption{Phase portraits for the SNU case. Left: $(B, \dot{B})$ plane, showing equilibrium points $B_{P_1}$ and $B_{P_2}$ (both centers). Right: $(a, \dot{a})$ plane, with the equilibrium point $a_{P_{SNU}}$ (center). Again, the solid black line indicates the possible flat universe, which separates the diagram into disjoint regions in terms of curvature. For these phase diagrams, we choose $\tilde{\beta}=1$.}
    \label{fig:nF}
\end{figure}

The qualitative behavior of the scale factor in the SNU universe is similar to that observed in the MU case. As seen in the right panel of figure~\ref{fig:nF}, when $\epsilon < 0$, the scale factor exhibits asymptotic behavior with a constant growth rate. For $\epsilon = 0$, the growth rate gradually decreases and tends to zero. In contrast, for $\epsilon > 0$, the scale factor reaches a finite maximum in finite time. In all cases, the system exhibits a bounce, characterized by a minimum in the scale factor followed by a period of accelerated expansion. Again, solutions with $\epsilon>0$ represent cyclic universes.

The behavior of the magnetic field, however, differs significantly depending on the spatial curvature (see the left panel of figure~\ref{fig:nF}). For $\epsilon = 0$, the field increases with constant acceleration, as indicated by the linear trend in the vector field. When $\epsilon < 0$, the acceleration of the magnetic field increases over time. In the case $\epsilon > 0$, the magnetic field reaches a maximum in finite time, similar to the behavior seen in the MU model, indicating a periodic behavior. 

\section{Concluding remarks}

The discussion presented in this work leads to a generalization of the Tolman–Ehrenfest procedure for spatial averaging, ensuring compatibility between nonlinear electrodynamics and a homogeneous, isotropic universe. Under suitable conditions, the usual approach is recovered, particularly for Lagrangians depending only on the invariant $F$. In this case, the spatial dependence of the invariant G plays no role. Moreover, because of the general form of anisotropic pressure and heat flux, the same compatibility can also be obtained whenever the Lagrangian is separable into a part depending only on $F$ and another depending only on $G$.

The qualitative analysis indicates the possibility of an accelerated expansion phase. Under different choices for the spatial curvature, it is possible to obtain universes that expand indefinitely for $\epsilon\leq0$, as well as universes with a maximum size that eventually recollapse once this upper limit is reached, in the case $\epsilon > 0$. The upper bound of the NEC (in terms of $B$) corresponds to an indeterminacy in the system constructed for the qMU. It is expected that this indeterminacy be removed by applying special techniques of the qualitative theory of singular fields.

Previously known results for the Magnetic Universe (MU) were recovered and complemented by a dynamical systems analysis. In this framework, both the NEC and the SEC can be violated, and solutions featuring bounces in the scale factor as well as cyclic universes are found. The opposite regime, approaching null fields with $F = 0$, was also analyzed. In this limit, the NEC is never violated, but the SEC still is, allowing for accelerated expansion. It is worth mentioning that these violations at the effective-fluid level signal that the background dynamics permits nonstandard behavior in general, but they do not by themselves establish whether the underlying field theory is free of ghosts or gradient instabilities. A definitive answer requires a perturbative analysis, an important and necessary follow-up to this qualitative study. Of course, in the weak-field regime, our quadratic model reduces smoothly to Maxwell theory and therefore inherits its stability.

Although the present work is qualitative, it suggests possible observational avenues to constrain the parameter space of the qMU. The strength and spectrum of primordial magnetic fields are bounded by CMB anisotropies \cite{Planck2016,Paoletti2022} and Faraday rotation measures \cite{Kahniashvili2009,Giovannini2018}, which indirectly constrain combinations of $\alpha$, $\beta$, and the initial field strength $B_0$. Furthermore, deviations in the early expansion history induced by NEC/SEC violations could leave imprints on primordial perturbations and the inflationary reheating era. A dedicated confrontation with CMB and large-scale-structure data, as well as bounds from magneto-genesis scenarios, is left for future work, but we outline the main tests here as guidance for follow-up studies.

Contrary to cosmology, where EM fields ought to be small, they are expected to be present and have very high intensity in the interior of stars. If this model is applied in the context of gravitational collapses, the peculiar pattern followed by the energy conditions may lead to unexpected results, possibly cyclic or bouncing behaviors inside the star. Another aim for further development is to analyze Lagrangians in which terms such as $\gamma F G$ are present, which will allow the extension to well-known Lagrangians like the Born–Infeld model. Finally, we also leave the study of these EM models in other background metrics and the analysis of perturbations for future work.

\section*{Acknowledgements}

The authors thank the support of CNPq (EB grant N.\ 305217/2022-4), CAPES (AGC grant N. 88887.666979/2022-00), and FAPERJ (MN is Emeritus Visiting Researcher fellow). FAF is supported by the PIBIC/UNIFEI Grant Program. We are indebted to the anonymous referees for their valuable comments, which greatly improved our manuscript, particularly through the suggestion to consider the Center Manifold Theorem.

\appendix

\section{The qualitative analysis of $\alpha<0$ case}\label{app:negative_alpha}

\begin{figure}
    \centering
    \includegraphics[width=0.3\linewidth]{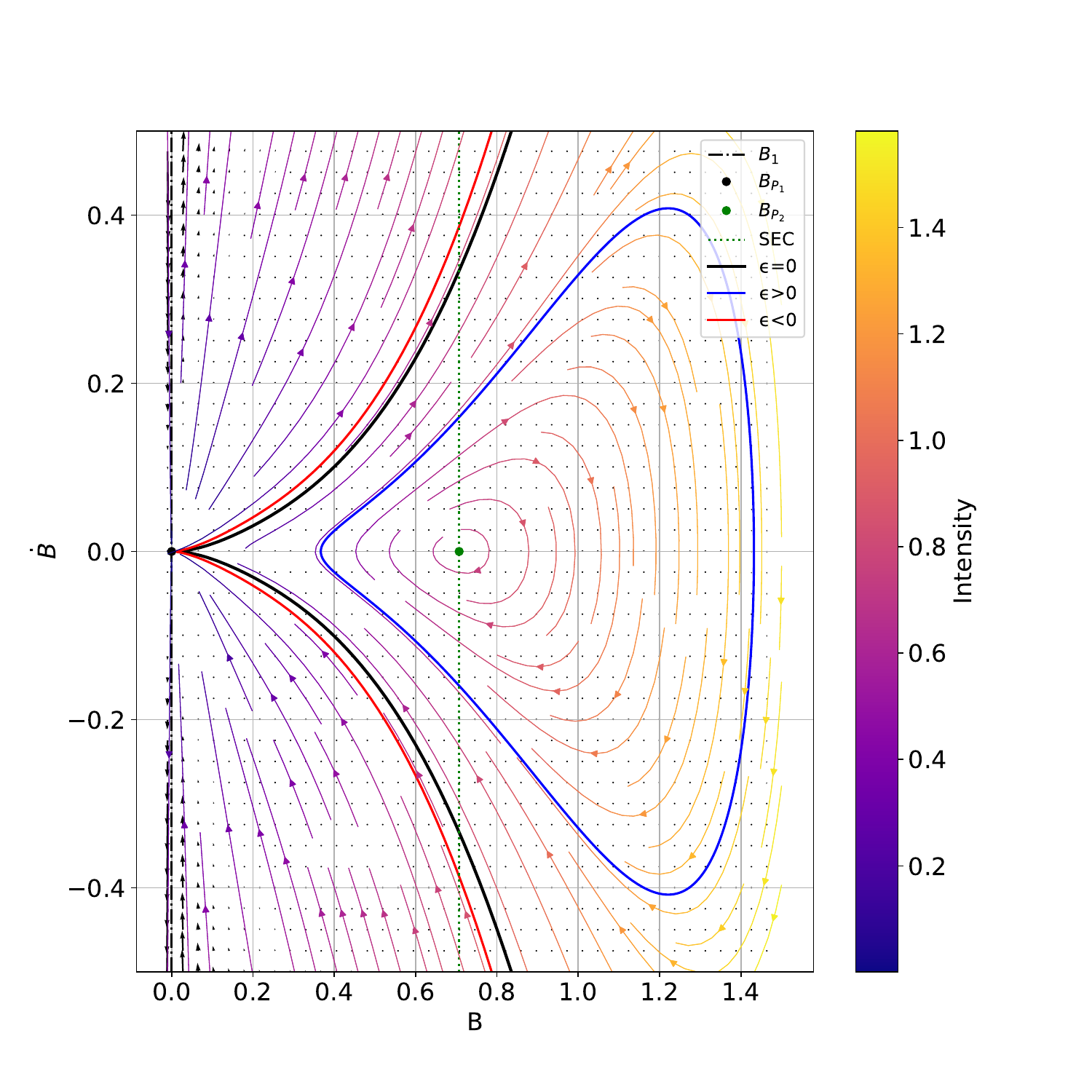}
    \includegraphics[width=0.3\linewidth]{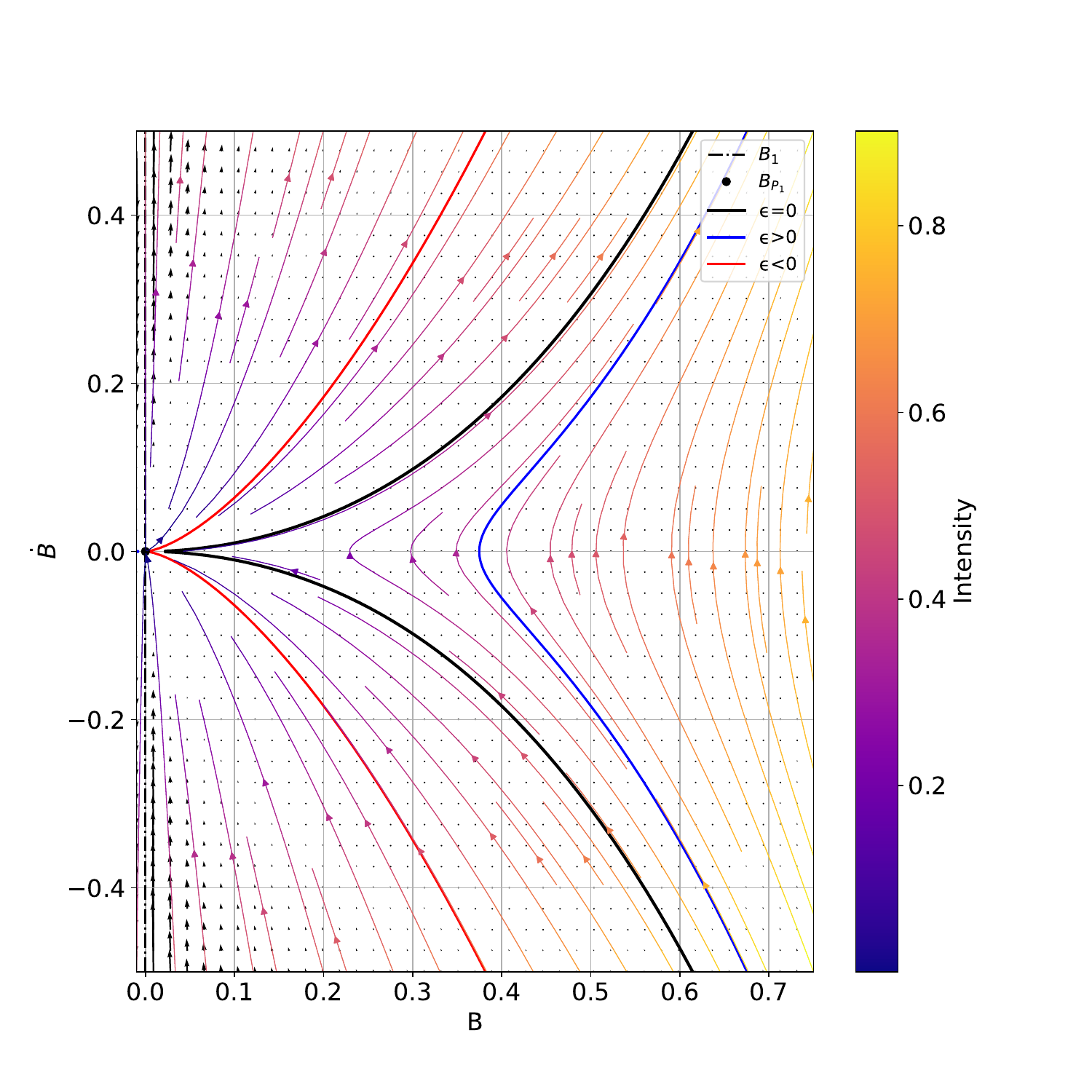}
    \includegraphics[width=0.3\linewidth]{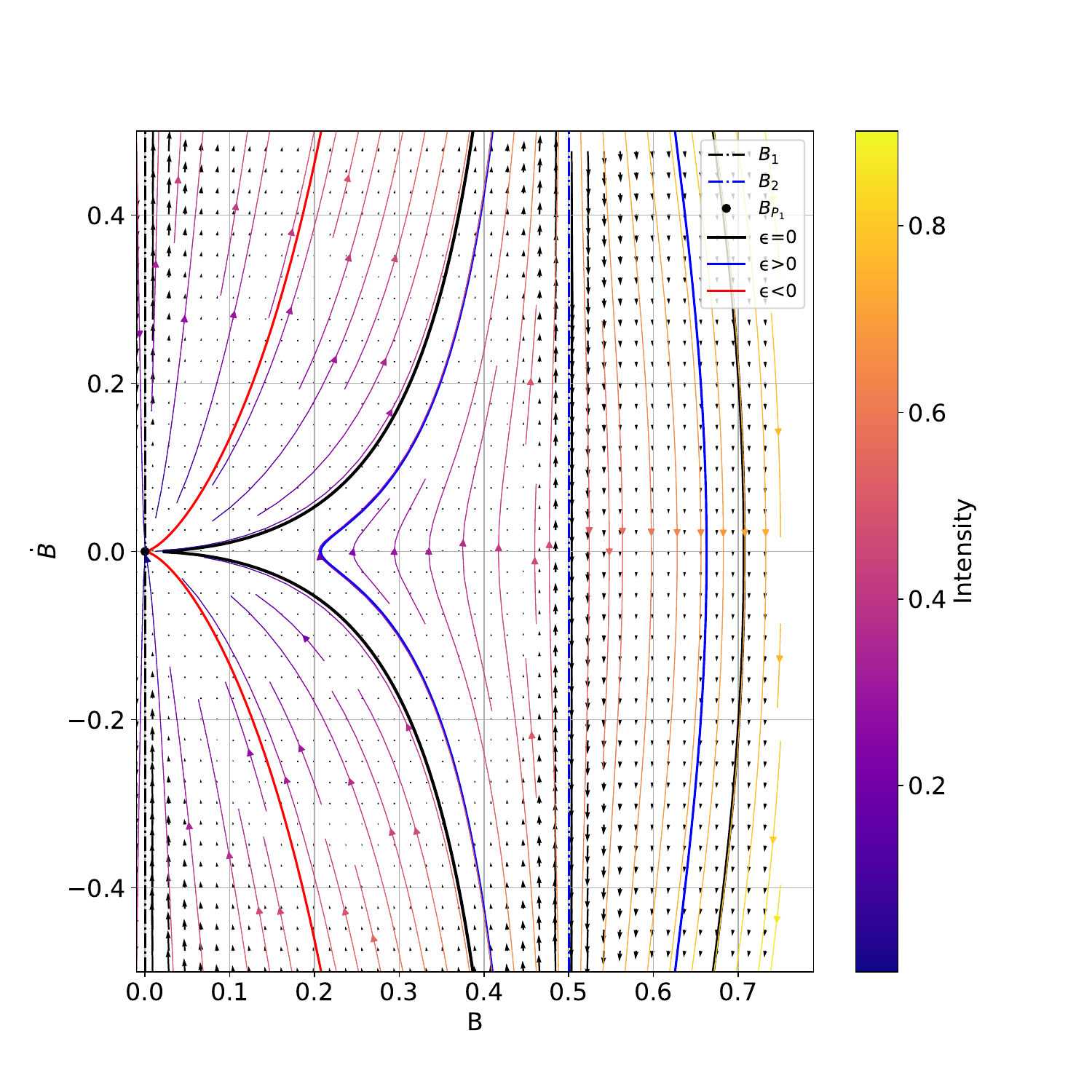}
    \caption{Phase Portraits when $\tilde\alpha<0$. Left: For Case I (we choose $w=-1$), the equilibrium points are the $B_{P_1}$ (``center'') and $B_{P_2}$ (center). Middle: For case II (we choose $w=0$), the only equilibrium is the origin (node). Right: For case III (we choose $w=1/2$), the only equilibrium is again the origin (node), but now there is a separatrix. For all phase diagrams, we choose $\tilde\alpha=-1$, and $\sigma=1/2$.}
    \label{fig:alphanegI}
\end{figure}

For the sake of completeness, we present here the possible phase space configurations when $\alpha<0$. The existence and behavior of the linearized dynamical system (\ref{sistemaquasimag}) is summarized in table \ref{tab:poss_quali_neg_alpha}, and the phase portraits corresponding to the three possible distinct cases are shown in figure \ref{fig:alphanegI}.
\begin{table}[ht]
    \centering
    \begin{tabular}{|c|c|c|c|c|}
    \hline
    Case & $w$ value & $B_{2}$ & $B_{P_2}$\\
    \hline
         I& $w<-\fracc{3}{4}$ & $\nexists$ & center\\
         II& $-\fracc{3}{4}<w<\fracc{1}{4}$ & $\nexists$ & $\nexists$\\
         III& $w>\fracc{1}{4}$ & $\exists$ & $\nexists$ \\ \hline
    \end{tabular}
    \caption{Existence of divergence line and equilibrium point (with its associated stability) for different values of $w$ and $\tilde\alpha<0$.}
    \label{tab:poss_quali_neg_alpha}
\end{table}

In Case I (left panel), for $\epsilon \leq 0$, the trajectories approach the equilibrium point $B_{P_1}$, indicating that the scale factor increases indefinitely. Concurrently, the magnetic field also grows, implying that the scale factor asymptotically approaches zero. For $\epsilon > 0$, the presence of both minima and maxima in the magnetic field solutions suggests that the scale factor oscillates between a minimum and a maximum value (a behavior that mirrors several other scenarios discussed in this work).

In Case II (middle panel), there is no equilibrium point apart from the origin, and all trajectories are unbounded. This corresponds to a scale factor that decreases toward zero, but only in the limit of infinite time. For $\epsilon > 0$, the presence of a minimum in the magnetic field indicates that the scale factor reaches an upper bound and starts to collapse after that.

In Case III (right panel), for small values of the magnetic field, an increase in $B$ corresponds to a decrease in the scale factor. However, beyond a critical value $B = B_2$, this trend reverses, and the scale factor begins to increase along with the magnetic field. This change in behavior occurs precisely at the minimum of the expression given in equation~(\ref{fatorequasimag}) for the chosen parameter values. A vertical line in the phase space marks a boundary beyond which the magnetic field cannot grow, which means that the scale factor reaches a non-singular minimum. For $B > B_2$, the magnetic field attains a maximum for all values of $\epsilon$, implying that the scale factor also reaches a corresponding maximum.

\section*{References}
\nocite{*}

\bibliography{ref.bib}

\end{document}